%% file: main.tex
\documentclass[letterpaper,twocolumn,10pt]{article}
\usepackage{usenix2019_v3}

\usepackage{cite}
\usepackage{amsmath,amssymb,amsfonts}
\usepackage{algorithmic}
\usepackage{array,multirow,graphicx}
\usepackage{textcomp}
\usepackage{listings}
\usepackage{xcolor}
\usepackage{caption}
\usepackage{listofitems}
\usepackage{adjustbox}

\usepackage{enumitem}
\usepackage{tikz}
\usepackage{pgfplots}
\usepackage{makecell}
\usepackage{pifont}
\usepackage{booktabs}
\usepackage{subcaption}
\usepackage{float}
\usepackage{color, colortbl}
\usepackage{nicematrix}
\usepackage{comment}
\usepackage{titlesec}
\usepackage{placeins}
\usepackage[normalem]{ulem}

\usepackage{url}

\usepackage{breakurl}
\usepackage{hyperref}
\hypersetup{colorlinks}

\definecolor{grey1}{gray}{0.91}
\def\code#1{\texttt{#1}}

\definecolor{dgray}{gray}{0.25}
\definecolor{strings}{RGB}{0,0,128}
\definecolor{backcolour}{rgb}{0.95,0.95,0.92}
\definecolor{keywords}{RGB}{127,0,85}
\definecolor{darkred}{RGB}{139,0,0}
\definecolor{darkyellow}{RGB}{204,204,0}
\definecolor{darkblue}{rgb}{0.0, 0.0, 0.55}
\definecolor{vividviolet}{rgb}{0.62, 0.0, 1.0}
\definecolor{fuchsia}{rgb}{1.0, 0.0, 1.0}
\definecolor{shockingpink}{rgb}{0.99, 0.06, 0.75}
\definecolor{darkBlue}{RGB}{0, 0, 205}
\definecolor{gray85}{gray}{0.85}
\usepackage[switch]{lineno}
\usepackage{xcolor}

\usepackage{graphicx}
\usepackage{hyperref}
\usepackage{caption}
\usepackage{subcaption}
\usepackage{soul} 
\usepackage{wrapfig}

\usepackage[most]{tcolorbox} 
\definecolor{findingbg}{RGB}{234,240,252}

\newcounter{findingcounter}
\newtcolorbox[auto counter]{finding}[1][]{
  colback=findingbg, 
  colframe=findingbg, 
  left=0mm, 
  right=0mm, 
  top=0mm, 
  bottom=0mm, 
  boxsep=1mm, 
  arc=0mm, 
  outer arc=0mm, 
  before upper={
  \refstepcounter{findingcounter}
  \textbf{Finding \thefindingcounter: }},
  #1
}

\begin{document}

\date{}

\title{\Large \bf Investigating Vulnerability Disclosures in Open-Source Software \\ Using Bug Bounty Reports and Security Advisories}


\author{
{\rm Jessy Ayala, Yu-Jye Tung, Joshua Garcia} \\ 
University of California, Irvine
} 

\maketitle

\begin{abstract}
In the world of open-source software (OSS), the number of known vulnerabilities has tremendously increased. 
The GitHub Advisory Database contains advisories for security risks in GitHub-hosted OSS projects. As of 09/25/2023, there are 197,609 unreviewed GitHub security advisories. 
Of those unreviewed, at least 63,852 are publicly documented vulnerabilities, potentially leaving many OSS projects vulnerable. 
Recently, bug bounty platforms have emerged to focus solely on providing bounties to help secure OSS. 
In this paper, we conduct an empirical study on 3,798 reviewed GitHub security advisories and 4,033 disclosed OSS bug bounty reports, a perspective that is currently understudied, because they contain comprehensive information about security incidents, e.g., the nature of vulnerabilities, their impact, and how they were resolved. 
We are the first to determine the explicit process describing how OSS vulnerabilities propagate from security advisories and bug bounty reports, which are the main intermediaries between vulnerability reporters, OSS maintainers, and dependent projects, to vulnerable OSS projects and entries in global vulnerability databases and possibly back.
This process uncovers how missing or delayed CVE assignments for OSS vulnerabilities result in projects, both in and out of OSS, not being notified of necessary security updates promptly and corresponding bottlenecks. 
Based on our findings, we provide suggestions, actionable items, and future research directions to help improve the security posture of OSS projects.
\end{abstract}


\input{content/introduction}
\input{content/background-and-rqs}
\input{content/methodology}
\input{content/empirical-study}
\input{content/discussion}
\input{content/related_work}
\input{content/conclusion}
\input{content/availability}
\bibliographystyle{plain}
\bibliography{main}
\input{content/appendix2}

\end{document}

%% file: content/introduction.tex
\section{Introduction}
With the explosion of open-source software (OSS), published OSS vulnerabilities have tremendously increased, reaching as high as 9,658 worldwide in 2020~\cite{statistaworldwide}. According to a recent report, high-risk vulnerabilities have increased by at least 42\% across all industry sectors since 2019, and most of them have an open-source component \cite{thenewstack}. It is no mystery that software vulnerability management is a challenge in the OSS ecosystem. Implementing vulnerability prevention mechanisms is likely the best way to build healthy, resilient systems \cite{ubuntublog}; however, it takes more than just upgrading a dependency to harden the security posture of an OSS project. 

Many reasons contribute to the exponential rise of OSS vulnerabilities, such as wide adoption of OSS by companies and in mobile applications \cite{venturebeat}, vulnerable dependencies in OSS software that are not patched by project maintainers promptly \cite{cybersecuritydive}, and supply chain attacks against package ecosystems like \texttt{npm} \cite{darkreading}. In June 2022, GitHub reported that there are more than 200 million active repositories \cite{usesignhouse}, a widespread attack surface for OSS vulnerabilities needing further investigation.

Most of the reasons stated above have been studied from various perspectives. 
For instance, Kula et al. \cite{kula2018developers} found that after investigating over 4,600 GitHub projects, 81.5\% kept outdated dependencies. 
Further, Liu et al. \cite{chengwei2022demystifying} perform an empirical study on vulnerability propagation and its evolution in the \texttt{npm} ecosystem and provide solutions for stakeholders to mitigate vulnerability impact. 
Prior work has found that maintainers tend to resolve security issues faster if there are associated CVEs \cite{buhlmann2022developers} and explore the role maintainers play in the OSS ecosystem \cite{akgul2023bug, alexopoulos2021vulnerability, maillart2017given}. 
Public security advisories and disclosed bug bounty reports can also help improve transparency about how prior vulnerabilities can be detected and fixed. 
In particular, prior work has studied the rate at which security reports are made into security advisories \cite{imtiaz2022open}, but not at the rate at which they are reviewed, i.e., for timely notification to vulnerable dependent client projects. 

Bug bounty platforms can also help secure OSS projects, where bug reporters can earn money as a ``bounty'' for reporting valid bugs and vulnerabilities. These platforms allow others to investigate publicly available code for OSS. Prior work on OSS bug bounty platforms have mainly focused on how to improve them \cite{atefi2023benefits,finifter2013empirical,ding2019ethical,elazari2018private,zhao2015empirical,walshe2022coordinated} and economic perspectives \cite{ruohonen2018bug,sridhar2021hacking,walshe2020empirical}, and industry perspectives~\cite{ahmed2024experience}, but not how OSS bug bounty reports are resolved so that they are properly routed to external databases.
Both security advisories and bug bounties play different roles in further securing OSS; however, \ul{using security advisories and bug bounty reports
to understand how OSS vulnerabilities are handled, their propagation throughout the OSS ecosystem, and their impact on OSS security posture 
is currently understudied}. 

\vspace{0.05cm}

In this paper, we conduct an empirical study on security advisories from GitHub Advisory Database (\textbf{GAD})
and OSS bug bounty reports. Upon scraping within constraints of APIs and publicly disclosed OSS bug bounty reports, we were able to gather 5,171 security advisories and 4,571 bug bounty reports before filtering. Our main contributions are as follows:

\vspace{-0.35cm}

\begin{itemize}
    \itemsep-0.4em 
    \item We are the first to determine the explicit process describing how OSS vulnerabilities propagate from security advisories and bug bounty reports to vulnerable projects and entries in global vulnerability databases and possibly back. 
    This process shows how missing or delayed CVE assignments for OSS vulnerabilities--regarding review turnaround times, time to reach NVD, reasons for not requesting a CVE, etc.--result in projects, in and out of OSS, not being notified of security updates promptly.
    \item We discover 47 CVE-assigned vulnerabilities that do not exist in the National Vulnerability Database (NVD) and reveal how popular OSS projects are still vulnerable to such CVEs. We inform MITRE of such CVEs, 12 of which were immediately added to the NVD as a result, and uncover that most are NVD-absent due to bottlenecks in the CVE Record Lifecycle. 
    \item We manually analyze 1,000 GitHub security advisories and OSS bug bounty reports to gather insight into how project maintainers handle OSS vulnerabilities and what it may take for such vulnerabilities to receive a CVE.
    \item We measure the usage of OSS vulnerability management features of 2,581 projects identified from GitHub security advisories and OSS bug bounty reports, which may indicate gaps during the vulnerability disclosure process.
    \item We provide an in-depth discussion, including suggestions and actionable items, which have corresponding insights to improve the security posture of OSS projects and have made relevant analytic data public~\cite{icse25repo}.
\end{itemize}


\vspace{-0.45cm}

%% file: content/background-and-rqs.tex
\section{\textbf{Background and Research Questions}}\label{ref:sec2}

To deepen our current understanding of OSS vulnerability management practices, we center our study on one of the largest OSS ecosystems, i.e., GitHub. 
An OSS ecosystem is made up of OSS project maintainers, vulnerability reporters, and client project developers. 
GAD~\cite{adv} and OSS-targeted bug bounty platforms~\cite{hunter,h1comm}  
help facilitate communication between the different actors in the GitHub OSS ecosystem. 
We do not include other OSS advisory databases, e.g., Snyk DB~\cite{snykdb}, because they do not directly map to GitHub projects.

\code{huntr} is an OSS bug bounty platform specifically for GitHub repositories.  
\code{huntr} pays vulnerability reporters to find vulnerabilities in GitHub repositories and project maintainers to fix them. 
By paying both vulnerability reporters and project maintainers, \code{huntr} encourages vulnerability reporters to report vulnerabilities and project maintainers to provide the fixes promptly. 
HackerOne, another popular bug bounty platform, provides vulnerability management services but does not pay vulnerability reporters or project maintainers.
However, HackerOne offers its services for free to OSS projects~\cite{hackeroneoss}.
We select active bug bounty platforms that are not (1) pay-to-use, e.g., YesWeHack, Bugcrowd; or (2) private, e.g., Synack, Intigriti, because our study requires mining disclosed bug bounty reports. 
We refrain from using bug bounty initiatives that are not security-focused, e.g., IssueHunt~\cite{issuehunt}, since our focus is security bug bounty reports, i.e., reports with assigned severities Low, Medium, High, or Critical.

On the other hand, GAD 
contains security advisories for publicly disclosed vulnerabilities. 
A GitHub security advisory is a publicly available announcement that discloses a vulnerability fix in a GitHub repository and alerts dependent client projects to update their dependencies.  
GAD sources advisories from other sources, e.g., National Vulnerability Database (NVD)~\cite{nistnvd} and FriendsOfPHP security advisories~\cite{fphp}, and those reported on GitHub directly. 

The advisories in GAD 
are divided into two groups: reviewed and unreviewed advisories. 
Reviewed advisories are reviewed by GitHub and are further tied to Dependabot where the advisories' dependent client projects are alerted~\cite{dpndbt}. 
Dependent client projects are not alerted of unreviewed advisories. 
Unreviewed advisories that stay unreviewed for a prolonged time can be dangerous for the dependent client projects since the dependent projects may not be aware of the vulnerability fixes, but advisories detailing such vulnerabilities are available for anyone to view.
Although reported vulnerabilities are not available to view until project maintainers provide the fixes, reported vulnerabilities that are not fixed promptly can risk rediscovery by bad actors. 
Reported vulnerabilities on either GAD, \code{huntr}, or HackerOne can be assigned a \emph{Common Vulnerabilities and Exposures} (CVE) identification.

To understand disclosure efficiency, i.e., factors affecting review rates and the flow to dependent projects for reports and advisories, that OSS vulnerabilities
face during the vulnerability review process,
we investigate the following RQs:   

\vspace{-0.4cm}

\begin{center}
\noindent\fbox{%
    \parbox{\columnwidth}{%
        \textbf{RQ1:} 
        To what extent is review turnaround time efficient for security advisories and bug bounty reports?
    }%
}
\end{center} 
\vspace{-0.5cm}
\begin{center}
\noindent\fbox{%
    \parbox{\columnwidth}{%
        \textbf{RQ2:}
        How efficiently do CVEs from security advisories and bug bounty reports get routed to NVD?
    }%
}
\end{center} 

\vspace{-0.15cm}

To obtain maximum exposure of an OSS bug bounty report, CVEs ensure that the vulnerability is routed to the NVD and further available to external advisory databases. Similarly, to obtain maximum exposure of a GitHub security advisory, a CVE must be created so that the vulnerability is routed to the NVD and is available to external advisory databases. Since GitHub security advisories belong to GAD, an alert will be generated regardless of CVE status once approved by GAD curators. 
To understand the reasons for successes, obstacles, and vulnerability-awareness prevention factors GitHub vulnerabilities face during the vulnerability disclosure process of security advisories and bug bounty reports, both with and without CVEs,
from GAD and bug bounty reports from \code{huntr} and HackerOne, 
we further investigate the following RQs:

\vspace{-0.45cm}

\begin{center}
\noindent\fbox{%
    \parbox{\columnwidth}{%
        \textbf{RQ3:} 
        What are the characteristics of bug bounty reports and security advisories with CVEs?
    }%
}
\end{center} 
\vspace{-0.3cm}
\begin{center}
\noindent\fbox{%
    \parbox{\columnwidth}{%
        \textbf{RQ4:} What prevents bug bounty reports and security advisories from obtaining CVEs?
    }%
}
\end{center}
\vspace{-0.3cm}
\begin{center}
\noindent\fbox{%
    \parbox{\columnwidth}{%
        \textbf{RQ5:}
        What do projects from security advisories and bug bounty reports tell us about gaps in vulnerability management features?
    }%
}
\end{center} 


%% file: content/methodology.tex
\begin{figure*}
    \centering
    \includegraphics[width=\linewidth,height=5.35cm]{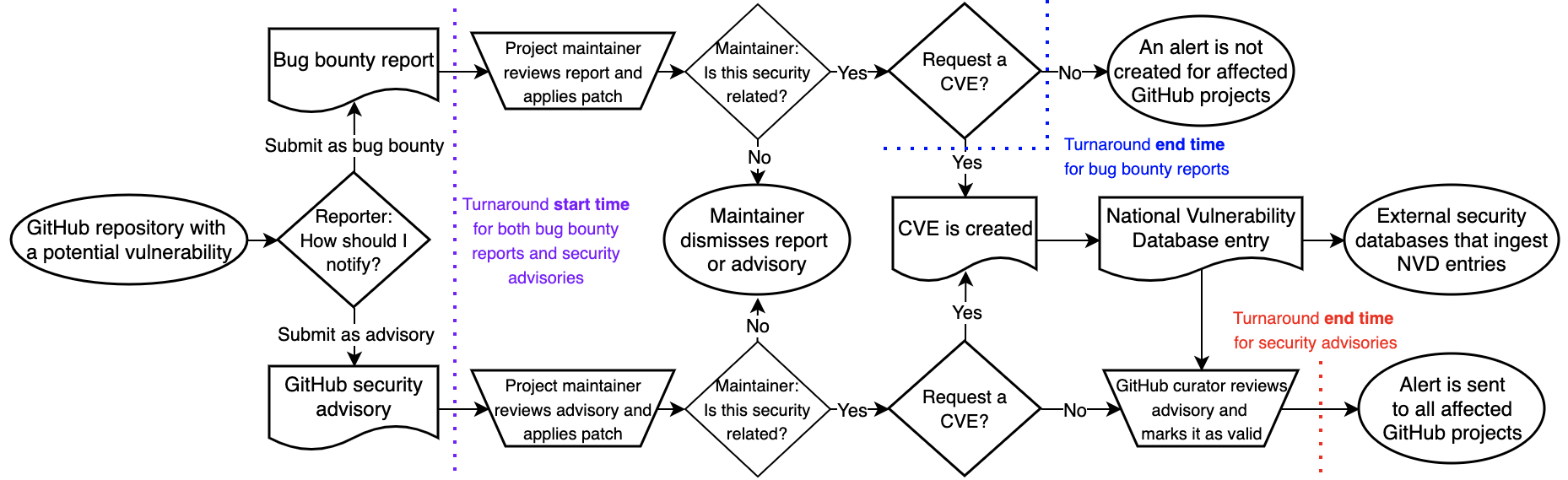}
    \caption{Report-and-Resolve Flows of an Open-source Vulnerability Using a Bug Bounty Program and GAD} 
    \label{fig:relationship}
\end{figure*} 

\vspace{-0.25cm}

\section{\textbf{Methodology}}\label{ref:sec3}

\vspace{-0.05cm}

We organize our study around two key components, security advisories and bug bounty reports. \hyperref[fig:relationship]{Figure 1} outlines the structure of a representative open-source vulnerability report-and-resolve process in the context of GitHub using GAD, 
a database containing open-source vulnerabilities, and either \code{huntr}, a bug bounty platform focused on GitHub projects, or HackerOne, a highly reputable bug bounty platform~\cite {h1popular}. 
Our study does not intend to draw correlations between bug bounty reports and security advisories, but instead explores and unveils OSS vulnerability management challenges, bottlenecks, and practices from these two understudied perspectives.

\vspace{0.1cm}

To obtain \autoref{fig:relationship}, we first analyzed existing bug-bounty reports and security advisories in our dataset to determine the flowlines between documents (e.g., bug bounty reports, created CVEs, and NVD entries), manual operations (e.g., project maintainer reviews) and decision points. 
We iteratively analyzed each report or advisory to see if a new flowline, manual operation, document, or decision point appeared until no additional flowchart elements arose. 
We further analyzed guidelines and lifecycles from \code{huntr} documentation~\cite{hunterguide} and security advisory documentation~\cite{githubadv}. Turnaround times are the report-to-resolve time frame, i.e., time to resolution, which includes the entire process of identification, triage, prioritization, patching, and final closure of the vulnerability. This is important because valid vulnerabilities should be patched promptly to avoid exploitation; further, it demonstrates a commitment to security and can protect project reputation, where time-efficient handling of vulnerabilities builds trust with users. Our study focuses on security advisories and bug bounty reports that reach a turnaround end time, 
as shown in \autoref{fig:relationship}, resulting in an alert sent to affected projects or no alert.

\vspace{-0.1cm}

\subsection{\textbf{Data Curation}}
As of 09/25/2023, there are 14,588 reviewed security advisories and 197,609 unreviewed security advisories on GAD, which date back to October 2017.
Of those unreviewed, at least 63,852 have an assigned CVE, potentially leaving many open-source projects vulnerable. Our collected GitHub security advisory sample is a subset of 14,588 reviewed security advisories, i.e., have both published and reviewed timestamps. 

We gather bug bounty reports from \code{huntr} and HackerOne. 
\code{huntr} does not provide a method of knowing how many bug bounty reports are publicly disclosed, nor a central list of projects with existing reports. 
We use its hacktivity page to collect projects from 09/01/2021 to 09/30/2023 and scrape reports per project; \code{huntr} reports retrieved date back to August 2019.
We gather HackerOne reports that include ``github.com'' in the report contents and have a known severity, ranging from February 2015 to January 2024, and manually label reports tied to GitHub projects since some results reference OSS but are closed-source, e.g., a report on internal software \cite{starbs}.

\vspace{-0.2cm}

\begin{table}[htbp]
\small 
\label{tab:emp0}
\begin{center}
    \begin{tabular}{ |p{0.07\textwidth}|p{0.16\textwidth}|p{0.16\textwidth}|} 
 \hline
 \textbf{Severity} & \textbf{Security Advisories} & \textbf{Bug Bounty Reports} \\ 
  \hline
 Low & 10.6\% (404/3,798) & 11.8\% (475/4,033) \\
  \hline
 Medium & 33.4\% (1,271/3,798) & 44.4\% (1,792/4,033) \\ 
 \hline
 High & 34.9\% (1,327/3,798) & 34.4\% (1,387/4,033) \\ 
 \hline 
 Critical & 20.9\% (796/3,798) & 8.9\% (357/4,033) \\
 \hline 
\end{tabular}
\end{center}
\caption{Security Advisories ($m$=3,798) and Bug Bounty (BB) Reports ($n$=4,033) Severities for GitHub Projects}
\end{table}

\vspace{-0.3cm}

Using all security advisories and bug bounty reports, we identified projects that are linked to GitHub repositories. We then query such repositories directly for the usage of configurable software vulnerability management features. This includes looking for a project vulnerability reporting policy, the ``Report a Vulnerability'' feature, and public security advisories. 
For GitHub, a repository's ``security policy" provides instructions on how to report a vulnerability, which is much more narrow than the meaning of security policy found in the research literature \cite{rsp1,rsp2,rsp3}. 
To avoid confusion with security policy's broader meaning, from here on forward, we will refer to GitHub's security policy as a vulnerability reporting policy. 
Repositories identified from GitHub security advisories and bug bounty reports have as many as 25,510,547 dependent projects~\cite{ljharbdeps} and 90,823 dependent packages~\cite{psfdeps}, indicating criticality to the software supply chain. In \hyperref[tab:emp0]{Table 1}, we show the severity breakdown of security advisories and bug bounty reports, with traceable GitHub source code, in our dataset and used for analysis (reflecting a similar consistency with the official NIST CVSS Severity Distribution Over Time~\cite{nvddist}).

\vspace{-0.1cm}

\subsection{\textbf{GitHub Security Advisories}}
We use the Puppeteer \cite{puppeteer} Node.JS library to scrape all queryable security advisory links marked as reviewed, i.e., by the GitHub Security Lab curation team, from each severity category from GAD. We scrape the 1,475 most recent security advisories from each category, except Low, which had 746 entries, resulting in 5,171 total advisory links; covering 1,987 GitHub projects. We use the GitHub API and \textsc{gh} CLI \cite{ghCLI} to gather metadata from each advisory, including its ID, CVE, severity, publish date, review date, and source URL

Upon further inspection, we find that the publish date of security advisories can be inconsistent. To mitigate this, we build a separate crawler, using \textsc{cURL} \cite{cURL} and \textsc{html2text} \cite{html2text}, to scrape security advisory publish dates. Similarly, we did this for correcting the \textit{nvd\_published\_at} column values, as metadata was inconsistent, i.e., metadata indicated 32.9\% of security advisories with a CVE were not cross-listed in the National Vulnerability Database (NVD), but it should be 0.8\% based on checking the NVD for CVEs marked as non-listed.  

\vspace{-0.1cm}

\subsection{\textbf{OSS Bug Bounty Reports}}


We use the Puppeteer \cite{puppeteer} Node.JS library, i.e., there is no \code{huntr} API, to scrape the 100 most recent disclosed bug bounty reports using all unique projects from the \code{huntr} hacktivity page \cite{huntr_hacktivity_page}, between September 2021 and September 2023, resulting in 3,181 reports. We leverage the HackerOne API to scrape all disclosed bug bounty reports that have a known severity and contain ``github.com'' 
in report contents to reduce the number of false positives, i.e., a report is not referring to an open-source project, resulting in 1,390 reports and 97 unique HackerOne project profiles that may correspond to zero or more GitHub repositories hosted per profile. Two researchers manually configure \textit{source\_code\_url} metadata of 900 HackerOne reports with corresponding GitHub project URLs to ensure analysis is only performed on bug bounty reports from GitHub source code. In total, bug bounty reports cover 701 GitHub projects.

For each report, we use Puppeteer \cite{puppeteer} to gather metadata from each identified bug bounty report, including its corresponding ID, CVE ID, severity, report date, disclosure date, and source code URL, \code{huntr}, or project profile owner, HackerOne. All bug bounty reports with a corresponding CVE ID were used to retrieve the date it was published in the NVD, if it exists, using \textsc{cURL} \cite{cURL} and \textsc{html2text} \cite{html2text}. 


\vspace{-0.1cm}

\subsection{\textbf{Usage of Software Vulnerability Management (SVM) Features} in OSS Projects}


For each GitHub project identified from security advisories and bug bounty reports, we gather their source code URL, whether they have a vulnerability reporting policy, whether they use the built-in ``Report a Vulnerability'' feature \cite{RaV_link} for creating GitHub security advisories, and whether security advisories are displayed on a project's security page. To do so, we retrieve the contents from webpages containing projects' vulnerability reporting policies and publicly visible security advisories with \textsc{cURL} \cite{cURL} and parse raw data with \textsc{html2text} \cite{html2text} to determine the existence of such software vulnerability management features. To further emphasize their importance and simplicity, we explain the purpose of and how to configure each SVM feature in \hyperref[tab:met4]{Table 2}.

\vspace{-0.2cm}

\begin{table}[htbp]\label{tab:met4}
\small
\begin{center}
    \begin{tabular}{ |p{0.185\columnwidth}|p{0.32\columnwidth}|p{0.34\columnwidth}|} 
 \hline
 \textbf{SVM \quad Feature} & \textbf{Purpose} & \textbf{How to Configure} \\ 
  \hline
 Vulnerability reporting policy & To give instructions for reporting security vulnerabilities. & Add a \code{SECURITY.md} file to the project's root, docs, or inside \code{.git}. \\ 
  \hline
 Public security advisories & To make it easier for the community to update package dependencies and research the impact of prior vulnerabilities. & In the \textit{Security} section of the sidebar, under \textit{Reporting}, click  \textit{Advisories}. From \textit{Security Advisories}, click the advisory to publish. \\ 
 \hline
 Private vulnerability reporting & To make it easier for privately reporting vulnerabilities directly to maintainers using a template. & In the \textit{Security} sidebar, click \textit{Code security and analysis} and enable \textit{Private vulnerability reporting}. \\ 
 \hline 
\end{tabular}
\end{center}
\vspace{-0.05cm}
\caption{Software Vulnerability Management (SVM) Feature Descriptions from GitHub \cite{github_code_security_page}}
\end{table}

\vspace{-1cm}



\subsection{\textbf{Data Analysis}}

\vspace{-0.05cm}

We use common data-science libraries implemented in Python to perform analysis using the information we collected from security advisories, bug bounty reports, and security vulnerability management features, e.g., vulnerability reporting policies. This includes metadata filtering, correcting data after additional web crawling, and calculating dataset statistics, such as turnaround time percentiles, and qualitative analyses \cite{icse25repo}. 

\vspace{-0.3cm}

%% file: content/empirical-study.tex
\section{\textbf{Empirical Study}}\label{ref:sec4}
\subsection{\textbf{RQ1: Review Turnaround Time for Security Advisories and Bug Bounty Reports}} \label{rq1}

In this section, we analyze review turnaround times, i.e., the duration between creation and closure, for security advisories and bug bounty reports. 
This analysis covers publicly disclosed GitHub security advisories and OSS bug bounty reports. 
We put our data in bins for review turnaround times, in \autoref{tab:turntable}, and is split into \textit{with CVEs} and \textit{without CVEs}, respectively. The full bin distribution and Mann-Whitney U test results deducing statistical significance are in our artifact \cite{icse25repo}.

\vspace{-0.2cm}

\begin{table}[htbp]    
    \small
    \begin{subtable}[h]{\columnwidth}
        \begin{tabular}{|p{0.28\columnwidth}|p{0.295\columnwidth}|p{0.275\columnwidth}|}  
        \hline
         \textbf{Turnaround Time} & \textbf{Sec. Advisories} & \textbf{BB Reports} \\ 
          \hline
         within a day & 58.3\% (1,995/3420) & 16.2\% (267/1,647) \\ 
         \hline
           within 2 weeks & 32.5\% (1,110/3420) & 33.2\% (546/1,647) \\ 
          \hline 
          within a month  & 1.6\% (56/3,420) & 14.9\% (245/1,647) \\ 
          \hline
          within 3 months & 1.0\% (35/3,420) & 20.5\% (337/1,647) \\
          \hline
          within 6 months & 1.3\% (45/3,420) & 3.2\% (45/1,647) \\ 
           \hline
        \end{tabular}
        \caption{\parbox{0.95\textwidth}{Review Turnaround Times (With CVEs) | $m$=3,420; $n$=1,647}} 
        \label{tab:emp2}
     \end{subtable}
     \vspace{0.2cm}
         \begin{subtable}[h]{\columnwidth}
    \begin{tabular}{|p{0.28\columnwidth}|p{0.295\columnwidth}|p{0.275\columnwidth}|}   
        \hline
         \textbf{Turnaround Time} & \textbf{Sec. Advisories} & \textbf{BB Reports} \\
  \hline
 within a day & 75.4\% (285/378) & 21.7\% (517/2,386) \\ 
  \hline
 within 2 weeks & 19.3\% (73/378) & 31.3\% (747/2,386) \\ 
 \hline 
 within a month  & 0.8\% (3/378) & 12.0\% (287/2,386) \\ 
 \hline
  within 3 months & 0.0\% (0/378) & 16.9\% (403/2,386) \\
  \hline
 within 6 months & 0.5\% (2/378) & 7.2\% (125/2,386) \\ 
 \hline
 \end{tabular}
       \caption{\parbox{0.95\textwidth}{Review Turnaround Times (Without CVEs) | $m$=378 $n$=2,386}} 
       \label{tab:emp3}
    \end{subtable}
     \caption{Turnaround Times for Reviewed Security Advisories ($m$) and Published Bug Bounty Reports ($n$)}
     \label{tab:turntable}
\end{table}

\vspace{-0.4cm}

Comparing values in \autoref{tab:emp2} against those in \autoref{tab:emp3}, we notice security advisories are resolved at a rate of 58\% to 75\% in a day, and only 19\% to 22\% for bug bounty reports. 
We perform Mann-Whitney U tests on turnaround times and find that \textit{Without CVEs} distributions are stochastically less than the \textit{With CVEs} distributions 
for both security advisories and bug bounties.
This reveals that bug bounty reports with CVEs take longer to resolve than those without a CVE, inconsistent with prior work ~\cite{buhlmann2022developers,farhang2020empirical}, as well as for security advisories, i.e., security advisories with CVEs are resolved slower than those without a CVE. 
Since CVEs are often associated with vulnerabilities that have a greater potential impact, these could be expected to be prioritized. 
In particular, Bühlmann et al.~\cite{buhlmann2022developers} found that GitHub security issue reports are resolved faster, especially if they have CVEs.
Similarly, Farhang et al.~\cite{farhang2020empirical} found that vendors are less likely to react with delay for CVEs with Android Git references.
However, we observe the opposite trend, contradicting the two prior works.
\begin{finding}\label{f1}
OSS bug bounty reports and GitHub security advisories with CVEs are resolved slower than those without a CVE. 
This finding contradicts prior work, which found that CVE-assigned vulnerabilities are resolved faster.
\end{finding}

\noindent\textit{\textbf{Diving Deeper Into Bug Bounty Reports}:} Given that only reviewed reports are in our dataset, i.e., unreviewed bug bounty reports are generally undisclosed in practice, we reach out to maintainers, whose projects have a history of CVE-assigned vulnerabilities, to gather additional context (with approval from our institution's IRB). 
We analyzed responses with by using inductive analysis \cite{induc2006thomas} and open coding~\cite{charmaz2014constructing}. Two researchers independently coded batches of eight responses at a time, resolving differences and updating the codebook after each batch, and synchronized codes via discussion.
In particular, two researchers analyzed the same 16 responses by engaging in open coding~\cite{charmaz2014constructing} and discussing the initial emerging themes, i.e., 3 out of 5 in \hyperref[tab:surveyresp]{Table 4}, meeting three times. Both researchers then independently coded the remaining 33 responses in batches of 7 and met frequently to discuss findings and reach consensus, resulting in 2 emerging codes after the initial analysis. Results are shown in \hyperref[tab:surveyresp]{Table 4}.

\vspace{-0.1cm}

\begin{table}[htbp]
\small
\begin{center}
    \begin{tabular}{ |p{0.2\columnwidth}|p{0.43\columnwidth}|p{0.21\columnwidth}|} 
 \hline
 \textbf{Reasoning} & \textbf{Representative quote} & \textbf{Distribution**} \\ 
  \hline
 Complexity and higher severity & ``Vulnerabilities that receive CVEs are often more complex and severe.'' & 38.8\% (19/49) \\ 
  \hline
 More formal process involved & ``Obtaining a CVE involves formal reporting/verification delays compared to handling vulnerabilities directly.'' & 34.7\% (17/49) \\ 
 \hline 
 Scrutiny and resource constraints & ``There is the anxiety of knowing a larger swath of the public will be examining your code [immediately]'' & 32.7\% (16/49) \\ 
 \hline 
 Coordination challenges & ``CVE-assigned vulnerabilities require coordination with researchers, affected users, and possibly other projects.'' & 14.3\% (7/49) \\ 
 \hline
 Not the case & ``We always treat the security issue as the most urgent issue rather than [dev] work.'' & 6.1\% (3/49) \\ 
 \hline 
\end{tabular}\label{tab:surveyresp}
\end{center}
\caption{Coded maintainer responses from an inquiry based on Finding~\ref{f1}. **Responses often contained multiple themes.
}
\end{table} 

\vspace{-0.4cm}

A majority of OSS maintainers (93.9\%) indicate having spent more time resolving CVE-assigned vulnerabilities than those without CVEs. Most cite ``complexity and higher severity'' (38.8\%), ``formal processes'' (34.7\%), and fear of public ``scrutiny and resource constraints'' (32.7\%); respondents also mention ``coordination challenges'' (14.3\%) with stakeholders.
On a larger scale, solo maintainers and smaller teams can be disproportionately affected by complexity, scrutiny, and coordination challenges, creating sustainability issues to secure smaller projects.
We encourage researchers to explore human-centered challenges and bottlenecks, e.g., via interview study, for primary stakeholders, i.e., OSS maintainers \ul{and} advisory database maintainers, concerning the trend where advisories and reports with CVEs are resolved slower since this goes against prior work and what one would expect in practice.

\begin{finding}\label{f2}
Based on 49 survey responses from OSS maintainers, 93.9\% resonate with our observation that CVE-assigned vulnerabilities are resolved slower than those without a CVE. Most respondents cite greater complexity/severity, formal processes, and fear of public scrutiny,
highlighting dynamics in how CVE-assigned vulnerabilities can slow down the resolution process for real-world OSS projects.
\end{finding}

\vspace{-0.15cm}

Inspired by Row 3 in \hyperref[tab:surveyresp]{Table 4} regarding resource constraints and prior work also considering bug bounty economics~\cite{ruohonen2018bug,walshe2020empirical}, e.g., bounty amounts, we also investigate monetary incentives.
In particular, two researchers manually labeled 1,000 randomly selected bug bounty reports (scrapers could not capture this information) with their corresponding bounty amount to calculate financial statistics. 
For disclosure bounties, reports with CVEs result in an average award of \$56.67; reports without CVEs result in an average award of \$54.89.
Other popular bug bounty platforms offer much higher bounties on average, e.g., the average bounty on HackerOne is \$1,000~\cite{h1report}. 
The much lower average bounty for OSS vulnerabilities indicates a lack of incentive for reporters to help develop security patches since OSS vulnerabilities pay much less than the average vulnerability on mainstream bug bounty platforms.
Fix bounties result in much smaller awards of \$4.95 for fixes with CVEs and \$9.18 without CVEs compared to disclosure bounties.
163 reports with CVEs offered no fix bounties; 71 reports without CVEs offered no fix bounties. 
We cannot replicate this process with security advisories because they do not have directly traceable monetary components.

\vspace{-0.15cm}

\begin{finding}\label{f22}
The lower award amount for fix bounties with CVEs than without and the higher rate of reports offering no fix bounty with CVEs than without them (2.30x greater) suggest a disincentive for hunters to develop security patches since maintainers have more fix bounty opportunities for reports without CVEs than those with CVEs. 
\end{finding}


\begin{figure}
    \includegraphics[width=\linewidth,height=4.25cm]{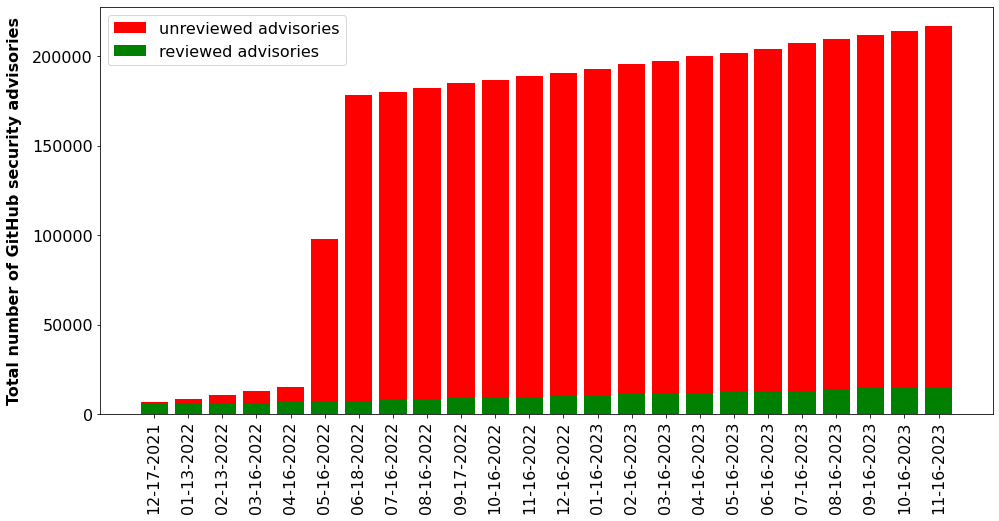} 
    \caption{Number of Reviewed and Unreviewed GitHub Security Advisories from December 2021 to November 2023}
    \label{fig:history}
\end{figure}

\noindent\textit{\textbf{Diving Deeper Into Security Advisories}:} \autoref{fig:history} shows the cumulative number of reviewed and unreviewed GitHub security advisories, green and red respectively, over two years, i.e., between December 2021 and November 2023. A sudden jump in unreviewed security advisories, concluding in June 2022, can be observed as a result of importing thousands of CVE entries from the National Vulnerability Database (NVD) ~\cite{gh2022nvd}. 
Using \autoref{tab:turntable} data, we find that the fastest review rate is 44 security advisories per day, found from reviews completed on 07/06/2023, and on average, 5.89 security advisories per day.
Based on the average rate, it would take 93.8 years to review all 201,687 advisories; however, it is unclear if GitHub is focusing on all advisories since some might be very old, and thus, they do not aim to review them anymore. We shift focus to security advisories introduced after major third-party imports. Using newer advisories, in July 2022, there were 836 reviewed and 847 unreviewed; as of November 2023, there were 7,571 reviewed and 30,720 unreviewed. 




\vspace{-0.15cm}

\begin{finding}[label=find:sec_adv_review_rate]\label{f3}
As of November 2023, 201,687 GitHub advisories remain unreviewed, largely due to a June 2022 import of CVEs from the NVD. At an average review rate of 5.89 advisories per day, clearing the backlog would take 93.8 years; focusing on newer advisories introduced after June 2022, it would take approximately 14.3 years to review all 30,720 of them. This highlights the challenge of scaling review efforts to address both legacy and recent advisories.
\end{finding}

\vspace{-0.3cm}

\subsection{\textbf{RQ2: Routing Vulnerabilities with CVEs to National Vulnerability Database (NVD)}} \label{streamling}

In this section, we consider GitHub security advisories and OSS bug bounty reports that have CVEs streamlined to the NVD, "the largest and most comprehensive database of known vulnerabilities" \cite{mendio}. 
This is also where GAD ingests entries with existing CVEs, as shown in \autoref{fig:relationship} and further evidenced by 99.3\% (214,092/215,566 on 11/02/2023) of advisories referencing the NVD.
Of the security advisories and bug bounty reports with assigned CVEs that exist in the NVD within our dataset, we find that 38.6\% (1,311/3,392) and 90.1\% (1,462/1,623), respectively, are new vulnerabilities. 

\vspace{-0.25cm}

\begin{table}[htbp]
    \small
        \begin{center}
        \begin{tabular}{|p{0.255\columnwidth}|p{0.3\columnwidth}|p{0.3\columnwidth}|} 
         \hline
        \textbf{NVD-Published} & \textbf{Sec. Advisories} & \textbf{BB Reports} \\ 
           \hline
           existed prior & 61.2\% (2,081/3392) & 9.9\% (161/1,623) \\ 
           \hline
           within a day & 31.4\% (1,065/3392) & 64.8\% (1,052/1623) \\ 
           \hline
           within 2 weeks & 5.7\% (193/3,392) & 18.4\% (299/1,623) \\ 
          \hline 
         within a month  & 0.6\% (21/3,392) & 2.0\% (33/1,623) \\ 
         \hline
         within 3 months & 0.4\% (14/3,392) & 1.0\% (17/1,623) \\
         \hline
       \end{tabular}
       \end{center}
       \caption{CVE Turnaround Times for Security Advisories ($m$=3,392) and Bug Bounty Reports ($n$=1,623) to NVD}
     \hfill
     \label{tab:turnnvd}
\end{table}

\vspace{-0.5cm}

From \autoref{tab:turnnvd}, we find that 0-day vulnerabilities from security advisories are transferred 1.78 times faster to the NVD than those from bug bounty reports within the week it is found. The faster integration of 0-day vulnerabilities from security advisories indicates prioritization to ensure prompt alerts are sent to affected projects. Based on this, maintainers should pay attention to security advisory alerts as they unveil older vulnerabilities and outdated dependencies, 61.4\% (2,081/3,392), and 0-day vulnerabilities, 38.6\% (1,311/3392).

Upon closer inspection of security advisories and bug bounty reports with a corresponding CVE, we find that there are approximately 1.85 times as many missing NVD entries coming from bug bounty reports than from security advisories in our dataset. In other words, it is less likely that a CVE-assigned security advisory has a missing NVD entry, such as ``Possible Denial of Service Vulnerability in Rack's header parsing'' \cite{adv1}, than a CVE-assigned bug bounty report, such as ``Failure to invalidate session after password change in \code{bigbluebutton/greenlight}'' \cite{anotherbb}, which does not exist in the security advisory database with query ``CVE-2022-36029'' \cite{adv}). This highlights a potential gap in the representation of security vulnerabilities in the NVD, raising concerns about its completeness and reliability. The oldest CVE absent from the NVD is CVE-2009-4123, which is 14 years old but can be found in the security advisory database~\cite{cve2009gh}. CVEs missing within the NVD, despite their presence in security advisories and bug bounty reports, suggest that there may be holes in the NVD data integration process; thereby, hindering the vulnerability disclosure process. 

\vspace{-0.1cm}

\begin{finding}\label{f4}
Even though security advisories contain new vulnerabilities less often, 38.6\% (1,311/3,392), than bug bounty reports, 90.1\% (1,462/1,623), they are streamlined to the NVD 1.85x faster. 
Further, CVEs from bug bounty reports are missing 1.78x more often in the NVD than CVEs from advisories, suggesting inefficiencies in verifying/documenting 0-day vulnerabilities from less formal channels.
\end{finding}
\vspace{-0.1cm}

We emailed MITRE to check on the 47 CVEs we found without an existing NVD entry. We learned that 4 were added after our data collection, 12 were added as a result of our email, and the other 31 have been acknowledged via email, 4 of which they were unaware of any public disclosure (we provided their corresponding disclosed bug bounty reports). They attribute the 27 missing CVEs to a delay in Step 6 shown on the CVE process page: ``once the minimum required data elements are included in the CVE Record, it is published to the CVE List by the responsible [CVE Numbering Authority] CNA'' ~\cite{cveprocess2}, e.g., a bug bounty provider. Though credible sources about a CVE Record are available, the completion of Step 6 is done by the responsible CNA, who participates voluntarily, rather than a central authority acting unilaterally. 

\vspace{-0.1cm}
\begin{finding}\label{f5}
We learn from MITRE that 66.0\% (31/47) of CVEs are NVD-absent due to a delay in CNAs publishing them to the CVE List. Such OSS CVEs are prevented from reaching external databases that ingest NVD entries, reducing vulnerability awareness.
\end{finding}
\vspace{-0.1cm}

\noindent\textit{\textbf{Case Study of CVE-2021-3902: }}
In the subset of CVEs in our dataset that are missing from the NVD, we conduct a case study on CVE-2021-3902, published on October 24, 2021. CVE-2021-3902 is of Critical severity and was detected in the \code{dompdf/dompdf} project, an HTML to PDF converter, which is vulnerable for software using versions under 2.0 due to improper restrictions of XML external entity reference for \code{svg} files. According to the CVE entry on the MITRE website, CVE-2021-3902 is Reserved, and ``This candidate has been reserved by an organization or individual that will use it when announcing a new security problem. When the candidate has been publicized, details for this candidate will be provided''~\cite{cve20213902}. 
Due to this vulnerability being reserved and not explained in the CVE entry, it does not exist in the NVD, which again, is taken by the GitHub security advisory curation team to be the primary source of CVEs, as evidenced by 99.3\% of security advisories referencing the NVD.

Within a PHP project, \code{composer.json} is used to specify common project properties, metadata, and dependencies. Using this information, we utilize the GitHub API to identify the top 100 repositories, based on the number of stars, that use \code{dompdf/dompdf} as a dependency in the \code{composer.json} file with the following query: \code{gh api --method=GET "search/code?q=dompdf/dompdf+filename:composer+} \code{extension:json+sort:stars"}. API limits are restricted to 100 results for code matching queries. 
We remove 11 inactive or unmaintained projects, as of 11/2023, and gather additional projects to conduct analysis~\cite{icse25repo}, as shown in \autoref{tab:cve20213902}. 


\begin{table}[htbp]
    \small
        \begin{center}
        \begin{tabular}{|p{0.2\columnwidth}|p{0.14\columnwidth}|p{0.16\columnwidth}|p{0.255\columnwidth}|}
         \hline
        \textbf{Max Version Supported} & \textbf{Release Year(s)} & \textbf{\# of Top Projects} & \textbf{Is Vulnerable to CVE-2021-3902} \\ 
           \hline
           0.* & 2013-20 & 48 & Yes \\ 
           \hline
           1.* & 2020-22 & 13 & Yes  \\ 
           \hline
           2.* & 2022-23 & 39 & No  \\ 
           \hline
       \end{tabular}
       \end{center}
         \caption{Top Projects Using \code{dompdf/dompdf} as a Dependency and Their Maximum Supported Versions ($k$=100)}
     \hfill
     \label{tab:cve20213902}
\end{table}

\vspace{-0.3cm}

From data in \autoref{tab:cve20213902}, we find that 61\% of the top 100 projects support a vulnerable version of \code{dompdf/dompdf} in their \code{composer.json} configuration file. Further, we found that \code{Attendize/Attendize}, which supports maximum version \code{dompdf/dompdf 0.8.6} and has 3,780 stars and 1,113 forks, is still vulnerable to CVE-2021-3902~\cite{attcommit}. It is surprising that a project as widely used as \code{dompdf/dompdf} can be vulnerable in many top projects, and the impact is compounded when the security advisory is not promptly created. 

These statistics also demonstrate the impact that interrupting the flow from Step 4, creating a CVE, to Step 5, becoming an NVD entry, from ~\autoref{fig:relationship}, can have on popular OSS GitHub projects. With such interruption, a GitHub security advisory is not created; thereby, alerts are prevented from being sent to all other affected projects. This is also the case for CVE-2021-3902, which is observed by zero search results when searching for ``CVE-2021-3902'' in GAD. We have submitted an advisory for this CVE to maximize exposure for affected projects and are waiting for it to be reviewed. Overall, CVE-2021-3902 highlights how the delay in transitioning from CVE creation to NVD entry can leave popular OSS projects exposed without the necessary security advisories. 


\begin{finding}\label{f6}
61\% of the top 100 repositories that use \code{dompdf} as a PHP dependency are vulnerable to CVE-2021-3902, a Critical CVE absent from the NVD. Since GitHub primarily pulls from the NVD for security advisories, 99.3\% of advisories as of November 2023, alerts for missing entries are prevented from being sent to all other affected projects. 
\end{finding}

\input{content/reworked-section43}

\subsection{\textbf{RQ5: Current Software Vulnerability Management (SVM) Feature Adoption in OSS}}\label{svmadopt}

In this section, we focus on the SVM feature usage in GitHub repositories found from each security advisory and bug bounty report. All bug bounty source code from reports are traceable to a GitHub repository: 707 GitHub projects. Not all security advisories contain a source code link within their metadata, in \textit{source\_code\_url}, so we use 77.6\% of security advisories with traceable source code: 1,982 GitHub projects. In \hyperref[tab:emp4]{Table 11}, we present usage statistics of vulnerability reporting policies, public security advisories, and private vulnerability reporting: \ul{2,581 unique GitHub projects total}. 

\begin{table}[htbp]
\small
\label{tab:emp4}
\begin{center}
    \begin{tabular}{|p{0.24\columnwidth}|p{0.305\columnwidth}|p{0.295\columnwidth}|} 
 \hline
 \textbf{SVM Feature Used} & \textbf{GitHub Projects (Sec. Advisories)} & \textbf{GitHub Projects (BB Reports)} \\ 
  \hline
  Vulnerability reporting policy & 52.7\% (1,045/1,982) & 76.5\% (541/707) \\ 
  \hline
 Public security advisories & 40.3\% (799/1,982) & 27.6\% (195/707)\\ 
 \hline
 Private vulnerability reporting & 23.1\% (457/1,982) & 29.0\% (205/707) \\ 
 \hline
\end{tabular}
\end{center}
\caption{SVM Feature Usage in Projects from Security Advisories ($m$=1,982) and Bug Bounty Reports ($n$=707)}
\end{table}

\vspace{0.1cm}

\noindent\textit{\textbf{Vulnerability Reporting Policies}:} As explained in \hyperref[tab:met4]{Table 2}, vulnerability reporting policies are important as they are meant to give instructions for reporting project-specific vulnerabilities. For instance, as stated in the \code{huntr} FAQ: ``your system needs a parseable e-mail in their \code{SECURITY.md} so your automation system can reach out to the maintainer'' \cite{hunter}. Looking at \hyperref[tab:emp4]{Table 11}, this statement reflects why a large percentage of projects, 76.5\% (541/707), of bug bounty reports have a vulnerability reporting policy, a reachable email is required for timely notification \cite{bbsp1,bbsp2,bbsp3,bbsp4,ayala2024mixed}. Further, 52.7\% (1,045/1,982) of projects discovered from GitHub security advisories have a respective vulnerability reporting policy. 


\begin{finding}\label{f11}
52.7\% (1,045/1,982) of projects from security advisories have a vulnerability reporting policy. Adding a \code{SECURITY.md} file benefits OSS projects as it opens the door for bug reporters that use vulnerability disclosure platforms, e.g., \code{huntr}, and other mechanisms, e.g., private email.
\end{finding}


\noindent\textit{\textbf{Public Security Advisories}:} Making known past security advisories public is meant to make it easier for the community to update package dependencies, research the impact of previously existing vulnerabilities, and adhere to responsible disclosure \cite{avastcite,vds2005}. \hyperref[tab:emp4]{Table 11} reveals that of the projects with reviewed security advisories, 59.7\% (1,183/1,982) have such advisories missing from their project page. This is concerning as project maintainers are given the option to publicize advisories if they are marked as reviewed in GAD. 

On the other hand, 72.4\% (512/707) of projects linked from bug bounty reports have security advisories missing on their project page. Upon closer inspection, we see from the first 250 advisories upon searching ``huntr'' in GAD---after discarding two duplicates---that there are no \code{huntr}-sourced advisories without a CVE. In our dataset, the corresponding date range, i.e., May 10, 2023, to Oct 1, 2023, contains 85 bug bounty reports without CVEs, meaning at least 85 \code{huntr} reports are missing. This outcome further validates a disclosure gap when bug bounty reports are missing a CVE.

\vspace{-0.1cm}

\begin{finding}\label{f12}
    63.4\% (1,636/2,581) of projects linked from security advisories and bug bounty reports do not have security advisories publicly displayed. This is a vulnerability disclosure gap as it hinders awareness of vulnerabilities and corresponding mitigation measures for affected projects.
\end{finding}



\noindent\textit{\textbf{Private Vulnerability Reporting}:} The ``Report a Vulnerability'' feature on GitHub is meant to make it easier for security researchers and bug finders to privately report vulnerabilities directly to project maintainers using a template form, which then become ``Unreviewed" entries in GAD. In \hyperref[tab:emp4]{Table 11}, there are roughly the same percentage of GitHub projects from \code{huntr} bug bounty reports with private vulnerability reporting enabled as those from security advisories, both under 30\%. The low adoption rate may suggest inefficiencies in reporting and addressing vulnerabilities within the OSS community. Further, projects that do not enable this feature may inadvertently discourage security researchers from responsibly disclosing vulnerabilities, which can negatively impact their security posture~\cite{disclosureblog,privdisclosure,gh2023private,ghissue0,ghissue1,ghissue2,ayala2024mixed}. 


\vspace{-0.1cm}

\begin{finding}\label{f13}
        76.1\% (1,963/2,581) of projects linked from security advisories and bug bounties have private vulnerability reporting disabled, which is concerning as it discourages researchers from responsibly disclosing vulnerabilities. Enabling this feature is trivial, but is surprisingly underused.
\end{finding}

\vspace{-0.4cm}

%% file: content/reworked-section43.tex

\subsection{\textbf{RQ3: Exploring Bug Bounty Reports and Security Advisories With CVEs}}\label{casestudy1}

In this section, we manually investigate 500 GitHub vulnerabilities, coming from bug bounty reports and security advisories, \underline{with} assigned CVEs (\autoref{tab:emp2}) and provide a descriptive analysis of such reports and advisories, i.e., as shown in \hyperref[tab:bbrcve]{Table 7} and \hyperref[tab:sacve]{Table 8}.
To do so, we leveraged qualitative methods designed to account for subjectivity by using inductive analysis \cite{induc2006thomas} and performed three rounds of open coding~\cite{charmaz2014constructing} for 125 out of 250 in each category to determine themes~\cite{braun2006using} in \hyperref[tab:bbrcve]{Table 7} and \hyperref[tab:sacve]{Table 8}. 
Two researchers independently coded batches of 25 reports and 25 advisories, resolved differences, and synchronized codes via discussion. Further, if a new code was adopted after discussion, researchers returned to re-code accordingly.

\vspace{0.05cm}

\noindent\textit{\textbf{Analysis of Bug Bounty Reports With CVEs}:} \label{bbranalysis}
In \hyperref[tab:bbrcve]{Table 7}, we present four distinct categories for CVE-assigned bug bounty reports, organized by their primary proof-of-concept (PoC) exploit.
We find that OSS project maintainers neither comment nor provide more than a simple ``thank you'' response in the majority, 55.0\% (44/80), of bug bounty reports containing a video or screenshots as the primary PoC. This suggests OSS project maintainers are more likely to request a CVE for bug bounty reports with visual evidence without further doubts, if any, than the other three categories presented: 36.0\% (27/75), 32.3\% (20/62), and 21.2\% (7/33), respectively. This favors vulnerability types, such as cross-site scripting, 20.4\% (51/250), that can be easily demonstrated visually.

\vspace{-0.2cm}

\begin{table}[htbp]
\small
\label{tab:bbrcve}
\begin{center}
    \begin{tabular}{|p{0.6\columnwidth}|p{0.29\columnwidth}|} 
 \hline
 \textbf{Primary PoC Type} & \textbf{BB Reports} \\ 
  \hline
 Video or screenshots & 32.0\% (80/250) \\
  \hline
 Step-by-step guide & 30.0\% (75/250) \\ 
 \hline
 Payload, e.g., \code{POST} request, script & 24.8\% (62/250) \\ 
 \hline 
 Custom input file, e.g., fuzz crash binary & 13.2\% (33/250) \\
 \hline 
\end{tabular}
\end{center}
\caption{PoC Type \% Distribution of OSS Bug Bounty Reports With Assigned CVEs ($n$=250)}
\end{table}

\vspace{-0.4cm}

In addition to categorizing CVE-assigned bug bounty reports, we examine conversations for any similarities and to gain further insight into the vulnerability disclosure process. We find that OSS project maintainers acknowledge reports that are easily reproducible \cite{bbrp1,bbrp2,bbrp3,ayala2024deep} 
or seen in a positive manner \cite{bbrpm1,bbrpm2,bbrpm3,ayala2024deep}. 
Further, a majority, 62.8\% (157/250), of bug bounty reports have a conversation, i.e., the reporter and project maintainer are active during review. This is essential for thorough bug bounty review, especially in cases where the reported vulnerability is not clear \cite{repro1,bbrp3,repro3,ayala2024deep,ayala2024mixed}. 
For instance, \code{huntr} disclosure requires maintainer permission to assign a CVE, unless the project is within the top 40\% of popular packages or repositories on GitHub \cite{bbrdp1,bbrdp2}.

\vspace{0.05cm}


\noindent\textit{\textbf{Analysis of Security Advisories With CVEs}:} 
In \hyperref[tab:sacve]{Table 8}, we present three categories of unique characteristics for CVE-assigned security advisories. 
We find that a majority have a public PoC, 47.2\% (118/250), or are directly imported from the NVD and exist in another advisory database, 32.8\% (82/250). The other advisories are submitted by users via GitHub private vulnerability reporting, 20.0\% (50/250). This suggests that security advisories not imported from the NVD are more likely to be assigned a CVE if they contain a PoC, 70.2\% (118/168); thus, reflecting that concrete evidence and trusted external validation are key drivers of CVE assignment. 

\vspace{-0.2cm}

\begin{table}[htbp]
\small
\label{tab:sacve}
\begin{center}
    \begin{tabular}{|p{0.6\columnwidth}|p{0.29\columnwidth}|} 
 \hline
 \textbf{Characteristic} & \textbf{Sec. Advisories} \\ 
  \hline
 Contains a PoC & 47.2\% (118/250) \\
  \hline
 NVD-imported \& exists in other database & 32.8\% (82/250) \\ 
 \hline
 User-reported with high-level description & 20.0\% (50/250) \\ 
 \hline 
\end{tabular}
\end{center}
\caption{Characteristics \% Distribution of GitHub Security Advisories With Assigned CVEs ($m$=250)}
\end{table}

\vspace{-0.4cm}

Security advisories with a PoC come from OSS bug bounty reports, 31.4\% (37/118), public issues, 17.8\% (21/118), and within the advisories themselves, 16.1\% (19/118). Those with public PoCs before publishing, 17.8\% (21/118), are concerning since it is bad practice to disclose vulnerabilities before they are patched~\cite{disclosureblog, privdisclosure,gh2023private}, especially if its exploit is included. Such projects were vulnerable for 367.7 days on average, ranging from 0 days to 1,677 days. The next category of advisories are imported from the NVD and exist in other security advisory databases, such as VulDB, suggesting that reviewers value sources beyond the NVD. The remaining advisories are user-reported with high-level vulnerability descriptions, which may lead to a higher risk of false positives or false negatives, and their small quantity suggests reviewers value advisories with technical detail or reputable, GitHub-external vulnerability advisory sources.

\vspace{0.1cm}

\noindent{\textbf{Summary}:} Having a reproducible exploit, references to external security advisories, or detailed descriptions to receive a CVE highlights the value of tangible evidence in the vulnerability assessment process; however, such standards may burden researchers and maintainers, particularly those with limited resources or expertise, creating barriers to getting CVEs for impactful vulnerabilities and an unintended disparity in how such vulnerabilities are represented post-triaging.
Further, an external security advisory acts as a form of delegation, beyond just reproducible exploits, that provides credibility, e.g., ~\cite{snykadv,ibexaadv}, by leveraging security community expertise; thereby, reducing investigative burden, e.g., reproducing complex but impactful vulnerabilities.
This approach, while rooted in credibility, only favors well-documented or externally validated vulnerabilities that receive CVEs, potentially leaving equally critical but harder-to-verify issues unaddressed.

\begin{finding}\label{f7}
86.8\% (217/250) of manually analyzed bug bounty reports with CVEs include video evidence, a step-by-step guide, or a specific payload.
Further, a majority of advisories not NVD-imported have a PoC, 70.2\% (118/168), where 17.8\% (21/118) were public before disclosure, leaving projects easily exploitable before a patch, for 368 days on average and ranging from 0 days to 6.3 years.
\end{finding}

\begin{finding}\label{f8}
Including a PoC can improve the chance that a security advisory or bug bounty report receives a CVE, but depending on the resources or expertise of researchers and maintainers, this may not always be possible. 
Besides PoC, an external security advisory can provide credibility for a security advisory or bug bounty report to receive a CVE. 
\end{finding}

\begin{table*}[htbp]
\fontsize{10}{12}\selectfont
\label{tab:emp8}
\begin{center}
    \begin{tabular}{ |p{0.18\textwidth}|p{0.63\textwidth}|p{0.12\textwidth}|} 
 \hline
 \textbf{Reasoning} & \textbf{Representative Quote} & \textbf{BB Reports} \\ 
  \hline
 No reason is stated & N/A; No assigned CVE is shown in the report, e.g., ~\cite{bbquote1} & 70.0\%(175/250) \\ 
  \hline
   Requested a CVE ID but did not hear back \quad \textit{**overlaps categories} & ``@maintainer Hi, It would be great if you publish a CVE for this, I wrote a \code{CodeQL} query to detect this pattern so anyone in open source community can use this to detect whether their repositories are vulnerable or not.''~\cite{bbquote3} & 13.2\%(33/250) \\
 \hline
     States that a fix or patch has been done &``I've discussed it with the team... I don't deem it necessary to push out a CVE notifying users to update to a new version.''~\cite{bbquote4} & 20.4\%(51/250) \\ 
 \hline 
  Poses little threat & ``You should try to write your reports by yourself. What have you learned after reporting this vulnerability? Just earn the bounty? The severity for this vulnerability is not High, it's Medium... You can check it on some reports or CVEs.'' \cite{bbquote6} & 9.6\% (24/250) \\
 \hline
\end{tabular}
\end{center}
\caption{Categories of Reasons Why a CVE is Not Submitted for OSS Bug Bounty Reports ($n$=250)}
\end{table*}

\begin{table*}[htbp]
\fontsize{10}{12}\selectfont
\label{tab:emp8p2}
\begin{center}
    \begin{tabular}{ |p{0.175\textwidth}|p{0.63\textwidth}|p{0.125\textwidth}|} 
 \hline
 \textbf{Characteristic} & \textbf{Representative Quote} & \textbf{Sec. Advisories} \\ 
  \hline
Recommends maintainers to upgrade & ``This incorrect behavior has been observed in real-world applications... ALL users of v0.7.1 and v0.7.2 [should] update to the latest version (v0.7.3), ASAP.''~\cite{adquote2} & 35.2\% (88/250) \\ 
  \hline
Applies patch \& assumes users will upgrade; no reason given & ``Jaeger UI is using the json-markup dependency to display span attributes and resources. This dependency is not sanitising keys of an object though, thus the KeyValuesTable is vulnerable to XSS.''~\cite{adquote1} & 30.8\% (77/250) \\ 
 \hline 
Limited impact or uncommon scenario & ``This impacted only the anonymous users themselves, and had no impact on logged in users... consider if this matters for your site.''~\cite{adquote3} & 24.4\% (61/250) \\ 
 \hline
Duplicate advisory & ``This advisory has been withdrawn because it is a duplicate of GHSA-[ID].''~\cite{adquote6} & 6.8\% (17/250) \\ 
\hline
Is still not patched & ``This high-severity vulnerability has been sitting in the package for months?... the age of the open issue this clearly reveals where security stands.''~\cite{adquote8} & 2.8\% (7/250) \\
\hline
\end{tabular}
\end{center}
\caption{Categories of Characteristics of GitHub Security Advisories Without a CVE ($m$=250)}
\end{table*}

\vspace{-0.3cm}

\subsection{\textbf{RQ4: Exploring Bug Bounty Reports and Security Advisories Without CVEs}}\label{casestudy1}

In this section, we manually investigate 500 GitHub vulnerabilities, coming from bug bounty reports and security advisories, \underline{without} assigned CVEs (\autoref{tab:emp3}) and provide a descriptive analysis to unveil potential reasons why such reports and advisories are missing CVEs, i.e., shown in \hyperref[tab:emp8]{Table 9} and \hyperref[tab:emp8p2]{Table 10}.
To do so, we again used qualitative methods designed to account for subjectivity with inductive analysis \cite{induc2006thomas} and performed three rounds of open coding~\cite{charmaz2014constructing} for 125 out of 250 in each category to determine themes~\cite{braun2006using} in \hyperref[tab:emp8]{Table 9} and \hyperref[tab:emp8p2]{Table 10}. 
Two researchers independently coded batches of 25 reports and 25 advisories, resolved differences, and synchronized codes via discussion. If a new code was adopted after discussion, researchers re-coded accordingly.

\vspace{0.05cm}

\noindent\textit{\textbf{Analysis of Bug Bounty Reports Without CVEs}:}
In \hyperref[tab:emp8]{Table 9}, we present reasons for bug bounty reports without assigned CVEs. Valid reasons for not filing a CVE request include ``poses little threat'' 9.6\% (24/250). However, we notice that a majority of reports without CVEs, 70.0\% (175/250), are missing reasons for not requesting one, i.e., the OSS maintainer does not state why a CVE was not requested. Further, ``patch makes exploit impossible or states that a fix has been done'', e.g., ``POC is too complicated''~\cite{bbquote8}, reflect 20.4\% (51/250) of analyzed reports. These are invalid reasons for neglecting to file a CVE since the vulnerability exists in prior versions. 
Depending on how well a bug bounty report PoC lines up with CVE criteria, it may still be qualified for one, leaving 90.4\% (226/250) of such reports analyzed without a valid reason for lacking a CVE. This is concerning because unassigned vulnerabilities pose real threats to systems, and their exclusion from CVE tracking hinders mitigation efforts~\cite{cveprocess}. Further, issues during review include a lack of communication (13.2\%). 



\vspace{0.05cm}

\noindent\textit{\textbf{Analysis of Security Advisories Without CVEs}:} 
In \hyperref[tab:emp8p2]{Table 10}, we present characteristics of GitHub security advisories without assigned CVEs and representative quotes that reflect such characteristics. Valid characteristics for not filing a CVE request for a security advisory include having ``limited impact'', 24.4\% (61/250), and being a ``duplicate of another advisory'', 31.2\% (78/250). However, the majority of advisories without an assigned CVE, 68.8\% (172/250), give a ``recommendation to upgrade'', contain a patch, and ``assumes users will upgrade'', or are ``still not patched''. Such advisories are concerning because they highlight many vulnerabilities with invalid reasons to not have a CVE, e.g., whether or not there is a patch has no bearing on CVE assignment, and are prevented from reaching external security advisory databases. 
Taking a closer look at CVE-absent security advisories without a valid reason, we find that 92.4\% (159/172) do not have a PoC, 68.0\% (117/172) do not reference an external security advisory, and 91.3\% (157/172) are user-reported. These statistics further reinforce metrics that support Finding \ref{f8}, e.g., security advisories should include a PoC to increase the chances of CVE assignment. The high percentage of user-reported advisories without a CVE raises questions about the reliability and technical depth of such advisories. 


\noindent{\textbf{Summary}:} To OSS maintainers, a CVE represents a more significant acknowledgment of a vulnerability than merely validating it, e.g., for a bounty. This could be attributed to varying levels of expertise or thresholds for what constitutes a vulnerability CVE-worthy and thereby, gatekeeping reports and advisories from CVE assignment.
For instance, instead of following up with reporters for a PoC or additional explanation after confirming security bugs as vulnerable, they will not give it a CVE until they are given a reproducible exploit---as indicated by 70.0\% (175/250) of non-CVE assigned bug bounty reports without conversation and the 68.8\% (172/250) of analyzed security advisories without valid characteristics for neglecting to request a CVE.
As a result, maintainers will prioritize deploying fixes but not associated vulnerabilities via the CVE process. 
This reflects an approach by maintainers who may not have the expertise to delve deeper into every reported vulnerability, but still address them through developing a security patch regardless of their interpreted importance.

\vspace{-0.2cm}

\begin{finding}\label{f9}
90.4\% (226/250) of bug bounty reports manually analyzed do not have a valid reason for neglecting to request a CVE. Such vulnerabilities are not routed to GAD, preventing alerts from being sent to affected projects. 
68.8\% (172/250) of analyzed security advisories do not have valid characteristics for neglecting to request a CVE and are prevented from reaching external databases that ingest NVD entries, further reducing vulnerability awareness.
\end{finding}

\vspace{-0.2cm}

\begin{finding}\label{f10}
A majority of bug bounty reports and security advisories analyzed do not have a valid reason or characteristic for not requesting a CVE, further indicating that CVEs are being withheld due to subjective or inconsistent thresholds by maintainers, rather than clear criteria.
Depending on OSS maintainers' levels of expertise or thresholds for what they deem as CVE-worthy, they can gatekeep security advisories or bug bounty reports from CVE assignments.
\end{finding}

\vspace{-0.3cm}

%% file: content/discussion.tex
\section{\textbf{Discussion}}\label{ref:sec5}

In this section, we discuss implications and future work, based on our findings, for OSS stakeholders and researchers,
followed by threats to external, internal, and construct validity in our study and the steps we have taken to mitigate them.

\vspace{-0.2cm}

\subsection{\textbf{Implications and Future Work}}


\textit{\textbf{For project maintainers}:}
We encourage OSS project maintainers who use GitHub to create security advisories for valid bug bounty reports, regardless of CVE status.
For CVE-assigned bug bounty reports, CVEs may still be NVD-absent (Findings~\ref{f4}, \ref{f5}, \ref{f6}), preventing them from being routed to GAD.
Bug bounty reports without CVEs do not exist in GAD.
We further encourage OSS project maintainers to
disclose the vulnerability on the respective project page to ensure OSS projects are alerted if they are affected~\cite{bbquote11}. 
Currently, as demonstrated in Finding \ref{f12}, many projects linked from security advisories and bug bounty reports do not publicly display security advisories.
Though we do not expect all CVEs in the NVD to have an advisory publicly displayed, it is best practice
to show relevant advisories potentially affecting other projects. 
We also advise OSS project maintainers to take advantage of available vulnerability management features, e.g., creating a vulnerability reporting policy and publicizing previously patched vulnerabilities, as these features are highly underutilized (Findings \ref{f11}, \ref{f12}, \ref{f13}).
Project maintainers can review a checklist of security expert-curated suggestions, e.g., Source Code Management Platform Configuration Best Practices by OpenSSF~\cite{scmopenssf}.
Lastly, we encourage project maintainers to review CVE-qualifying criteria before deciding not to file a request since valid vulnerabilities are absent from security advisories, as demonstrated by Findings \ref{f9} and \ref{f10}, 
preventing alerts from being sent to affected projects. 

Tooling should be developed to address automation challenges to make vulnerability management for reports and advisories operate more efficiently, effectively, and standardized (Findings \ref{f7}, \ref{f8}, \ref{f9}, \ref{f10}), i.e., automated tooling should assist project maintainers with (1) determining the extent to which a vulnerability from a security advisory or bug bounty report affects their project; and (2) aid in vulnerability identification and repair for related vulnerabilities that may exist elsewhere in the project. 
Research challenges that make creating such automation difficult include assisted or automated vulnerability scoring and problems with minimizing regressions when automatically generating security fixes, both of which are currently understudied, as far as we are aware.
\vspace{0.05cm}

\noindent\textit{\textbf{For code hosting platforms}:} 
OSS code hosting platforms should consider enabling security-related features by default to encourage secure code development, effective vulnerability management, and transparency in the OSS ecosystem. For instance, private vulnerability reporting is currently disabled by default in GitHub. 
As shown in Finding \ref{f13}, many projects, 76.1\% (1,963/2,581), keep this default setting.
We suggest it is enabled by default for all GitHub projects to avoid unattended security issues from being publicly disclosed, especially for such projects that do not have a clear point-of-contact listed. 
The disable option should remain since some projects have an existing vulnerability reporting process based on prior complaints in early 2023, e.g., ``My team already has external ways to report those sorts of issues, so we don't use Github for them'' ~\cite{disableconvo}. 
Further, we suggest that security advisories in GitHub projects are public by default to maximize vulnerability exposure, which is generally important in OSS~\cite{disclosureblog}. This way, the OSS community and dependent projects are aware of the risk of adopting previously vulnerable projects; thereby, allowing a broader awareness of vulnerabilities within the OSS community and downstream dependent projects to encourage transparency and accountability (general OSS principles).

\vspace{0.1cm}

\noindent\textit{\textbf{For vulnerability database maintainers}:} We suggest vulnerability database maintainers conduct routine check-ins and host public meetings regarding vulnerability maintenance for communal involvement. 
For instance, we communicated with MITRE about bottlenecks in the CVE-to-NVD streamlining process, and they determined that CNAs are a major bottleneck with CVE publishing as demonstrated from Finding \ref{f5}; thus, we suggest MITRE refines this process to be more proactive, where CNAs are periodically contacted for a CVE update. 
Although we recognize the issue underlying this finding is difficult to address, e.g., CNAs participate voluntarily, we hope our study will support community discourse where reasonable remedies can be collaboratively planned and implemented, such as developing techniques to automate and scale vulnerability management tasks and study their effectiveness.
Vulnerability database maintainers can work towards this goal by flagging entries that have been stale or have not made their way through the disclosure process, making note of particular CNAs or organizations that come up frequently.

Additional approaches should be explored to identify specific bottlenecks that can be further addressed to improve the efficacy of (1) the CVE Record Lifecycle; and (2) accurately assessing the overwhelming amount of unreviewed security advisories.
Potential research challenges include determining how human factors influence the extent to which the CVE Record Lifecycle is delayed, e.g., how the burden on CNAs can be alleviated, and identifying what it means for a vulnerability to be considered valid, especially when PoCs are complex in nature or limited information is known, so that GitHub Advisory Database curators can speed up turnaround times for advisories' awareness and dependent client projects are not overwhelmed with dependency upgrade notifications~\cite{ayala2024deep}.

\vspace{0.05cm}

\noindent\textit{\textbf{For researchers}:}
Beyond the research challenges described prior, we suggest researchers explore approaches to help reduce overhead during the review process, e.g., the current review rate cannot keep up with the number of unreviewed advisories (Finding \ref{f3});
and explore further the trend where security advisories and bug bounty reports with CVEs are resolved slower since this is not in line with prior work and goes against what one would expect in practice (Finding \ref{f1}).
As a result, tools can be developed to support OSS stakeholders in conducting effective vulnerability management to reduce technical challenges (Finding \ref{f2}); further, provide incentives for improving OSS project security posture (Findings \ref{f2} and \ref{f22}).
Lastly, we encourage researchers to investigate how the presence, or absence, of vulnerability reporting policies impacts the security resilience of OSS projects, e.g., by studying how vulnerability-contributing commits~\cite{vccs,vccfinder} are handled across OSS projects, that do or do not have a vulnerability reporting policy, along with policy contents to gather insights about important policy components. 
Researchers can also use our dataset and participate in community-driven OSS initiatives, e.g., OpenSSF has active working groups including curated datasets that can be used for studies~\cite{openssfwg}, to help advance techniques in the vulnerability lifecycle and communication between stakeholders; thereby, working closely with project maintainers and vulnerability database maintainers to understand real-world challenges and gain practical insights.

\vspace{-0.3cm}

\subsection{\textbf{Threats to Validity}}

\noindent\textbf{\textit{External validity}} reflects the extent to which results from our study can be generalized. 
One threat to the external validity of our study is whether our findings and insights can be generalized to other OSS ecosystems.
To mitigate this threat, we center our study on one of the largest OSS ecosystems, i.e., GitHub. 
While GitHub provides a representative foundation for studying OSS practices, we recognize that platform-specific features, e.g., private vulnerability reporting and GAD, and community dynamics may influence our findings.
Another threat to the external validity of our study is whether our findings and insights can be generalized to other OSS security advisories and OSS bug bounty reports. 
To mitigate this threat, we maximize our dataset beyond API limits for scraping. 
For GitHub security advisories, to maximize our dataset, we gathered at most 1,475 advisories per severity category, resulting in 5,171 unique security advisories before filtering and 1,982 unique GitHub projects. 
To address overfitting to advisory sources, we note that GitHub security advisories are not solely reported by GitHub stakeholders, but include advisories from many databases such as RustSec Security and Python Packaging Advisories~\cite{globaladv}.

For OSS bug bounty reports, we pull from \code{huntr} and HackerOne. 
We collect bug bounty report data from these two bug bounty platforms and believe such data may be generalizable because (1) \code{huntr} is the only active bug bounty platform focused on GitHub projects and (2) HackerOne is the overall most widely-used bug bounty platform and offers its services for free to OSS projects~\cite{hackeroneoss}, including those hosted on GitHub~.
We analyze 3,133 \code{huntr} bug bounty reports, covering 567 unique GitHub projects, ranging from September 2021 to September 2023. 
Further, popular platforms might have more resources and streamlined processes, while less prominent platforms might struggle with resource constraints, affecting their response and resolution times; to mitigate this, we ensure that the analysis includes data from both popular and OSS-specific platforms, to get a comprehensive view of the OSS vulnerability landscape.
To address overfitting to GitHub projects per bug bounty platform, we also analyze 900 HackerOne bug bounty reports, covering 144 unique GitHub projects, ranging from February 2015 to January 2024. 

\vspace{0.1cm}

\noindent\textbf{\textit{Internal validity}} reflects the extent to which a cause-and-effect link cannot be explained by alternative factors. A threat to internal validity is our ability to retrieve accurate and complete metadata associated with security advisories and bug bounty reports. 
To mitigate this threat, we built additional scrapers to expand our dataset beyond API constraints for holistic coverage. 
Since we used on APIs for collecting data, we found some metadata to be inconsistent, such as \textit{github\_published\_at} timestamps in advisories. 
Further, we found some aspects of bug bounty reports to be unscrapable when crawling them; thus, we correct metadata by building additional scrapers and merging them, such as \textit{corrected\_date}. 

\vspace{0.05cm}

\noindent\textbf{\textit{Construct validity}} addresses the extent to which the measurements employed in our study accurately and meaningfully represent the concepts we aim to assess. 
A threat to construct validity for our study is manually labeling security advisories and bug bounty reports. 
We mitigate this threat by looking beyond surface-level advisories and reports, and explore documentation linked to their respective writeups, e.g., GitHub issues. 
Moreover, two researchers independently review advisories and reports, and resolve discrepancies via discussion.

\vspace{-0.2cm}

%% file: content/related_work.tex
\section{\textbf{Related Work}}\label{ref:sec6}

\vspace{-0.1cm}

We divide previous work on OSS into four categories: 
(1) studies on the role open-source project maintainers have played,
(2) studies on the role vulnerability reporters have played,
(3) studies centered on bug bounty platforms, and
(4) studies centered on other OSS aspects.
We discuss each category and
conclude with distinctions between prior work and our work.

\vspace{0.05cm}

\noindent\textit{\textbf{Open-source project maintainer roles}:}
Researchers have studied the role open-source project maintainers have played in the OSS ecosystem~\cite{buhlmann2022developers,wang2019detecting,farhang2020empirical,ayala2024deep,ayala2024mixed}. 
Ayala et al.~\cite{ayala2024deep,ayala2024mixed} studied the features that OSS project maintainers want for handling bug bounty reports~\cite{ayala2024deep} and their perspective on vulnerability management features~\cite{ayala2024mixed}.
Wang et al.~\cite{wang2019detecting} found that some OSS project maintainers will secretly patch vulnerabilities, and proposed an approach to detect such
security patches. Their work indicates that hiding security issues, or ``security by obscurity,'' does not make their patches significantly more difficult for attackers to find, but may instead negatively affect projects as developers are unaware of critical updates.
Buhlmann et al.~\cite{buhlmann2022developers} found that maintainers tend to
resolve security issues faster if they have associated CVEs.
Farhang et al.~\cite{farhang2020empirical} made the same observation that security issues are resolved faster with associated CVEs for the Android ecosystem.
These works disagree with our finding that reports and advisories without CVEs have faster resolution.

\vspace{0.05cm}

\noindent\textit{\textbf{Vulnerability reporter roles}:}
Researchers have also studied the role vulnerability reporters have played 
in the OSS ecosystem~\cite{akgul2023bug,alexopoulos2021vulnerability,maillart2017given}. 
Alexopoulos et al.~\cite{alexopoulos2021vulnerability} and Ruohonen et al.~\cite{ruohonen2018bug} found that a small group of vulnerability reporters account for a large number of the reports. 
However, Alexopoulos et al. also found that vulnerability reporters who reported to a project
for the first time accounted for a majority of reports. 
Maillart et al.~\cite{maillart2017given} found that vulnerability reporters specialize in finding specific vulnerabilities. These findings affirm the value of OSS bug bounty platforms as ``given enough eyeballs, all bugs are shallow''~\cite{raymond1999cathedral}.
Akgul et al.~\cite{akgul2023bug} found that vulnerability reporters' main dissatisfaction when reporting is poor communication from the project maintainers. Shafigh et al.~\cite{shafigh2021bugbounty} created a taxonomy for out-of-scope HackerOne bug bounty reports. The \code{huntr} bug bounty platform, which also gives bounties to OSS project maintainers for fixing reported vulnerabilities, requires maintainers to actively engage with vulnerability reporters. Our work includes reports from \code{huntr} and HackerOne to understand the current state of OSS.

\vspace{0.04cm}

\noindent\textit{\textbf{Bug bounty platforms}:}
Previous work have also centered on understanding bug bounty platforms. 
Luna et al.~\cite{luna2019productivity} 
investigated the productivity of vulnerability reporters across different bug bounty platforms.
Many researchers have investigated how to improve bug bounty platforms~\cite{atefi2023benefits,finifter2013empirical,ding2019ethical,elazari2018private,zhao2015empirical,zhao2017devising,walshe2022coordinated} and some have also studied bug bounty platform economics~\cite{ruohonen2018bug,sridhar2021hacking,walshe2020empirical}. 
Further, bug bounty program policies have been evaluated to determine how to promote ethical hacking for hunters and responsible disclosure~\cite{elazari2018policy, laszka2018engagement}.

\vspace{0.04cm}

\noindent\textit{\textbf{Other relevant aspects of open-source software}:}
There are also work that investigated other OSS aspects: the lack of platform security features, e.g., security policies and security-related workflows in GitHub projects~\cite{ayala2023empirical,ayala2024mixed}, the rate at which security reports are made 
into public security advisories~\cite{imtiaz2022open}, the reluctance of developers to update outdated dependencies~\cite{kula2018developers}, and suggestions to 
improve developers' trust in dependency bots used to alert when a dependency is reported via security advisory~\cite{he2023automating}.  

\vspace{0.04cm}

\noindent\textit{\textbf{Our study}:}
\ul{No previous work on OSS performed an empirical study on reports in public security advisories and bug bounty platforms}, which are 
the main intermediaries between vulnerability reporters, OSS maintainers, and dependent client projects; therefore, both public 
security advisories and bug bounty platforms play an important co-existing 
role in shaping the OSS ecosystem. 
As our research has shown, key insights on the state of software vulnerability management can be discovered using GitHub security advisories and OSS bug bounty reports. 
These insights not only shed light on current practices, but also highlight the gaps and challenges in collaboration, triaging, and communication between 
stakeholders for proper disclosure and raising awareness of vulnerabilities. 

\vspace{-0.2cm}

%% file: content/conclusion.tex
\section{\textbf{Conclusion}}\label{ref:sec7}

\vspace{-0.05cm}

In this paper, we conduct an empirical study to investigate OSS vulnerability disclosure practices from the perspectives of OSS projects with disclosed bug bounty reports and reviewed GitHub security advisories. 
Based on our findings, we suggest that OSS platforms should consider enabling easily configurable security features by default, e.g., private vulnerability reporting, OSS maintainers should make an effort to communicate with bug bounty reporters since most review bug bounty reports without discussion, and vulnerability database maintainers use indicators for stale CVEs after a specified time to identify bottlenecks for CVEs ``stuck'' in the CVE Record Lifecycle.
Future work should investigate and develop techniques towards automated vulnerability management tooling for OSS maintainers and automated CVE analysis tooling for vulnerability database maintainers, i.e., to improve the efficacy of the CVE Record Lifecycle.

%% file: content/availability.tex

\section{Ethics considerations}\label{sec:ethics}

Our institution’s ethics review board (IRB) approved the portion of our study in which we reached out to OSS project maintainers, i.e., to fill out a survey, in our dataset.
To comply with GitHub's terms of service~\cite{githubTOS} and recent studies that consult OSS project maintainers~\cite{utz2022privacy,ayala2024deep,ayala2024mixed}, we only reached out to maintainers who have public contact information advertised as reachable to the general public, e.g., in their profile introduction markdown, or through a hosted website, e.g., in their personal homepage.


\section{Open science}

Our analyses and dataset are available publicly online~\cite{icse25repo}.

%% file: content/appendix2.tex
\appendix

\section{New Security Advisories Post-NVD Import}

\vspace{-0.4cm}

\begin{figure}[h]
    \centering
    \includegraphics[width=\linewidth,height=4.25cm]{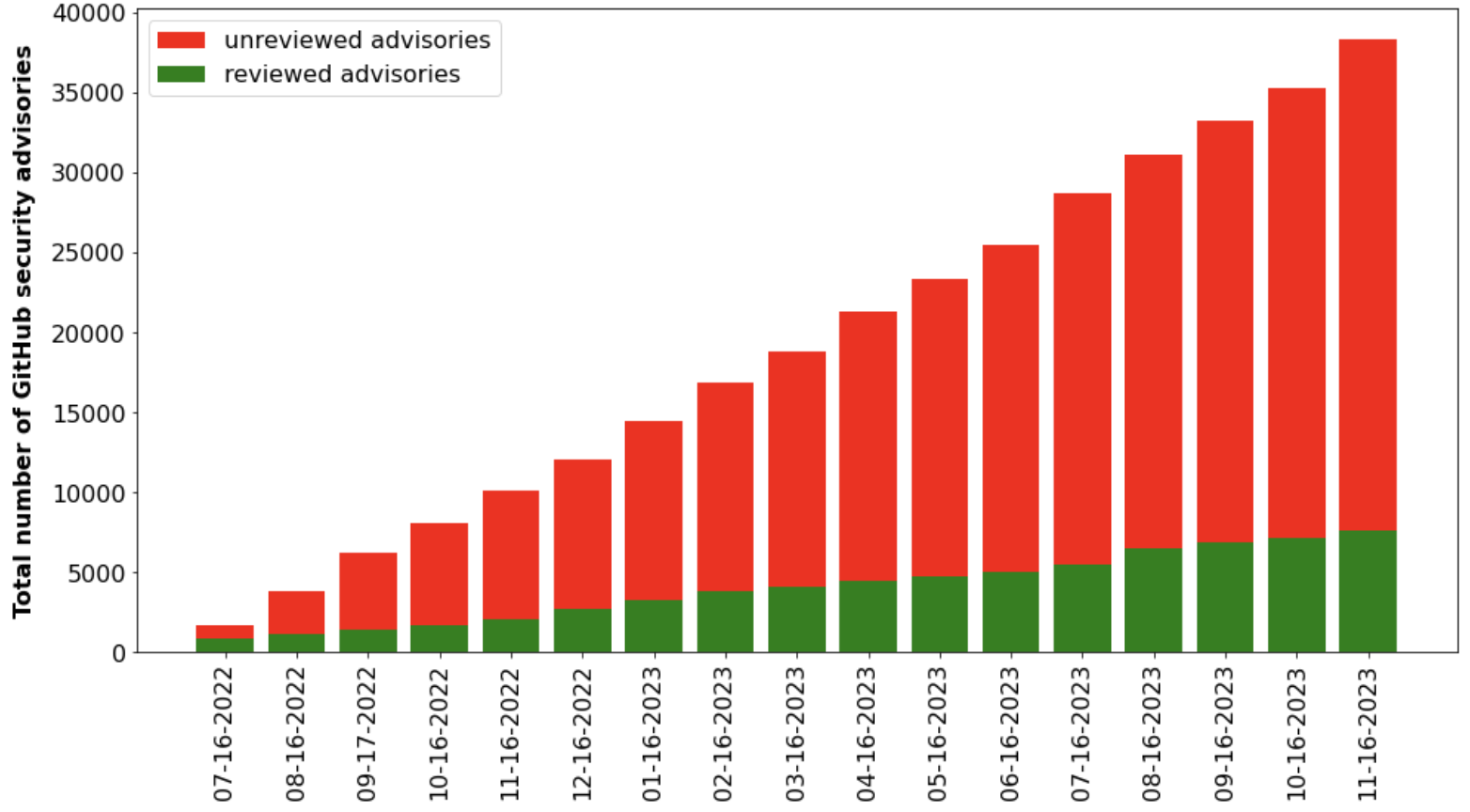} 
    \caption{Number of \underline{New} Reviewed and Unreviewed GitHub Security Advisories from June 2022 (i.e., After the Large Number of NVD Imported Entries) to November 2023}
\end{figure}

\vspace{-0.3cm}




%% file: main.bbl
\begin{thebibliography}{100}

\bibitem{icse25repo}
Anonymized repository for open science.
\newblock \url{https://anonymous.4open.science/r/usenix2025-vulnerability-disclosures-repo-3D71/README.md}, 2025.

\bibitem{bbrpm2}
Affan Ahmed.
\newblock File upload type validation error in unilogies/bumsys.
\newblock \url{https://huntr.com/bounties/b5e9c578-1a33-4745-bf6b-e7cdb89793f7/}, 2023.

\bibitem{ahmed2024experience}
Ali Ahmed, Ho~Cheung~Brian Lee, and Amit Deokar.
\newblock Experience and efficiency in vulnerability resolution on bug bounty platforms (research-in-progress).
\newblock {\em Proceedings of the 19th Pre-ICIS Workshop on Information Security and Privacy}, 2024.

\bibitem{akgul2023bug}
Omer Akgul, Taha Eghtesad, Amit Elazari, Omprakash Gnawali, Jens Grossklags, Michelle~L Mazurek, Daniel Votipka, and Aron Laszka.
\newblock Bug hunters’ perspectives on the challenges and benefits of the bug bounty ecosystem.
\newblock In {\em 32nd USENIX Security Symposium (USENIX Security)}, volume 2301, 2023.

\bibitem{rsp3}
Ahmad Al-Omari, Omar El-Gayar, and Amit Deokar.
\newblock Security policy compliance: User acceptance perspective.
\newblock In {\em 2012 45th Hawaii International Conference on System Sciences}, pages 3317--3326, 2012.

\bibitem{alexopoulos2021vulnerability}
Nikolaos Alexopoulos, Andrew Meneely, Dorian Arnouts, and Max M{\"u}hlh{\"a}user.
\newblock Who are vulnerability reporters? a large-scale empirical study on floss.
\newblock In {\em Proceedings of the 15th ACM/IEEE international symposium on empirical software engineering and measurement (ESEM)}, pages 1--12, 2021.

\bibitem{bbsp3}
Melbin~Mathew Antony.
\newblock Cross-site scripting (xss) - reflected in area17/twill.
\newblock \url{https://huntr.com/bounties/89ef143b-4829-41db-b31b-75c1e03a300f/}, 2021.

\bibitem{vds2005}
A.~Arora and R.~Telang.
\newblock Economics of software vulnerability disclosure.
\newblock {\em IEEE Security \& Privacy}, 3(01):20--25, January 2005.

\bibitem{repro3}
asura n.
\newblock Cross-site request forgery (csrf) in kevinpapst/kimai2.
\newblock \url{https://huntr.com/bounties/5fa3098a-ba02-45e0-af56-645e34dbc691/}, 2021.

\bibitem{atefi2023benefits}
Soodeh Atefi, Amutheezan Sivagnanam, Afiya Ayman, Jens Grossklags, and Aron Laszka.
\newblock The benefits of vulnerability discovery and bug bounty programs: Case studies of chromium and firefox.
\newblock In {\em Proceedings of the ACM Web Conference 2023}, pages 2209--2219, 2023.

\bibitem{ayala2023empirical}
Jessy Ayala and Joshua Garcia.
\newblock An empirical study on workflows and security policies in popular github repositoriess.
\newblock In {\em 2023 IEEE/ACM 1st International Workshop on Software Vulnerability Management (SVM)}, pages 6--9, 2023.

\bibitem{ayala2024deep}
Jessy Ayala, Steven Ngo, and Joshua Garcia.
\newblock A deep dive into how open-source project maintainers review and resolve bug bounty reports, 2024.

\bibitem{ayala2024mixed}
Jessy Ayala, Yu-Jye Tung, and Joshua Garcia.
\newblock A mixed-methods study of open-source software maintainers on vulnerability management and platform security features, 2024.

\bibitem{bbquote6}
Devendra Bhatla.
\newblock Cross-site request forgery (csrf) in splitbrain/dokuwiki.
\newblock \url{https://huntr.com/bounties/e20fc1c1-3b42-4900-9983-7afa36cb681c}, 2021.

\bibitem{braun2006using}
Virginia Braun and Victoria Clarke.
\newblock Using thematic analysis in psychology.
\newblock {\em Qualitative research in psychology}, 3(2):77--101, 2006.

\bibitem{buhlmann2022developers}
Noah B{\"u}hlmann and Mohammad Ghafari.
\newblock How do developers deal with security issue reports on github?
\newblock In {\em Proceedings of the 37th ACM/SIGAPP Symposium on Applied Computing}, pages 1580--1589, 2022.

\bibitem{ghissue2}
Thomas Castronovo.
\newblock Security issue \#480.
\newblock \url{https://github.com/Romanitho/Winget-AutoUpdate/issues/480}, 2023.

\bibitem{charmaz2014constructing}
Kathy Charmaz.
\newblock Constructing grounded theory.
\newblock 2014.

\bibitem{attcommit}
Johanna Cherry.
\newblock Attendize: A free and open-source event management and ticket selling application.
\newblock \url{https://github.com/Attendize/Attendize/blob/9289acbab1583898fd85aeee66c7b613d8971deb/composer.json}, 2016.

\bibitem{bbrpm1}
Axel Chong.
\newblock Improper access control in bookstackapp/bookstack.
\newblock \url{https://huntr.com/bounties/135f2d7d-ab0b-4351-99b9-889efac46fca/}, 2021.

\bibitem{bbrpm3}
Axel Chong.
\newblock Improper access control in snipe/snipe-it.
\newblock \url{https://huntr.com/bounties/efdf2ead-f9d1-4767-9f02-d11f762d15e7/}, 2022.

\bibitem{bbquote4}
Axel Chong.
\newblock Rce in wordnet browser in nltk/nltk.
\newblock \url{https://huntr.com/bounties/cd3957f0-2c9c-416d-bc3a-190a5b7ce4a6}, 2022.

\bibitem{usesignhouse}
Ch~Daniel.
\newblock Github users and growth statistics: How many repos are there? (2023).
\newblock \url{https://www.usesignhouse.com/blog/github-stats}, 2023.

\bibitem{adquote6}
René de~Sain.
\newblock Duplicate advisory: Grafana stored cross-site scripting vulnerability.
\newblock \url{https://github.com/advisories/GHSA-3cgw-hfw7-wc7j}, 2023.

\bibitem{puppeteer}
Chrome DevTools.
\newblock Puppeteer.
\newblock \url{https://pptr.dev/}, 2017.

\bibitem{ding2019ethical}
Aaron~Yi Ding, Gianluca~Limon De~Jesus, and Marijn Janssen.
\newblock Ethical hacking for boosting iot vulnerability management: A first look into bug bounty programs and responsible disclosure.
\newblock In {\em Proceedings of the Eighth International Conference on Telecommunications and Remote Sensing}, pages 49--55, 2019.

\bibitem{elazari2018policy}
Ami Elazari.
\newblock Private ordering shaping cybersecurity policy: The case of bug bounties.
\newblock 2018.

\bibitem{elazari2018private}
Amit Elazari.
\newblock Private ordering shaping cybersecurity policy: The case of bug bounties.
\newblock {\em Rewired: Cybersecurity Governance}, pages 231--264, 2018.

\bibitem{farhang2020empirical}
Sadegh Farhang, Mehmet~Bahadir Kirdan, Aron Laszka, and Jens Grossklags.
\newblock An empirical study of android security bulletins in different vendors.
\newblock In {\em Proceedings of The Web Conference 2020}, pages 3063--3069, 2020.

\bibitem{finifter2013empirical}
Matthew Finifter, Devdatta Akhawe, and David Wagner.
\newblock An empirical study of vulnerability rewards programs.
\newblock In {\em 22nd USENIX Security Symposium (USENIX Security 13)}, pages 273--288, 2013.

\bibitem{thenewstack}
Charlotte Freeman.
\newblock Open source vulnerabilities are still a challenge for developers.
\newblock \url{https://thenewstack.io/open-source-vulnerabilities-still-a-challenge-for-developers}, 2023.

\bibitem{fphp}
FriendsOfPHP.
\newblock Php security advisories.
\newblock \url{https://github.com/FriendsOfPHP/security-advisories}, 2014.

\bibitem{disclosureblog}
Nancy Gariché.
\newblock Coordinated vulnerability disclosure (cvd) for open source projects.
\newblock \url{https://github.blog/2022-02-09-coordinated-vulnerability-disclosure-cvd-open-source-projects/}, 2022.

\bibitem{githubTOS}
GitHub.
\newblock Github terms of service.
\newblock \url{https://docs.github.com/en/site-policy/acceptable-use-policies/github-acceptable-use-policies}.

\bibitem{globaladv}
GitHub.
\newblock About global security advisories.
\newblock \url{https://docs.github.com/en/code-security/security-advisories/working-with-global-security-advisories-from-the-github-advisory-database/about-global-security-advisories}, 2017.

\bibitem{github_code_security_page}
GitHub.
\newblock Code security documentation.
\newblock \url{https://docs.github.com/en/code-security}, 2017.

\bibitem{RaV_link}
GitHub.
\newblock Configuring private vulnerability reporting for a repository.
\newblock \url{https://docs.github.com/en/code-security/security-advisories/working-with-repository-security-advisories/configuring-private-vulnerability-reporting-for-a-repository}, 2017.

\bibitem{adv}
GitHub.
\newblock Github security advisory database.
\newblock \url{https://github.com/advisories}, 2017.

\bibitem{githubadv}
GitHub.
\newblock Publishing a repository security advisory.
\newblock \url{https://docs.github.com/en/code-security/security-advisories/working-with-repository-security-advisories/publishing-a-repository-security-advisory}, 2017.

\bibitem{dpndbt}
GitHub.
\newblock Dependabot: Automated dependency updates built into github.
\newblock \url{https://github.com/dependabot}, 2019.

\bibitem{ghCLI}
GitHub.
\newblock Github command line.
\newblock \url{https://cli.github.com/}, 2020.

\bibitem{gh2022nvd}
GitHub.
\newblock All historical nvd advisories are now listed on github.
\newblock \url{https://github.blog/changelog/2022-06-08-all-historical-nvd-advisories-are-now-listed-on-github}, 2022.

\bibitem{gh2023private}
GitHub.
\newblock Private vulnerability reporting now generally available.
\newblock \url{https://github.blog/2023-04-19-private-vulnerability-reporting-now-generally-available/}, 2023.

\bibitem{ghissue0}
Julian Gonggrijp.
\newblock Issue templates \#4278.
\newblock \url{https://github.com/jashkenas/backbone/pull/4278}, 2023.

\bibitem{ubuntublog}
Massimiliano Gori.
\newblock Open source security: from prevention to recovery.
\newblock \url{https://ubuntu.com/blog/open-source-and-cybersecurity-from-prevention-to-recovery}, 2022.

\bibitem{adquote2}
Adam Greig.
\newblock Miscompilation in cortex-m-rt 0.7.1 and 0.7.2.
\newblock \url{https://github.com/advisories/GHSA-xw5j-gv2g-mjm2}, 2023.

\bibitem{adquote1}
Sven Grossmann.
\newblock A stored xss in jaeger ui might allow an attacker who controls a trace to perform arbitrary jaeger queries.
\newblock \url{https://github.com/advisories/GHSA-2w8w-qhg4-f78j}, 2023.

\bibitem{h1comm}
HackerOne.
\newblock Hackerone community edition.
\newblock \url{https://www.hackerone.com/company/open-source-community}, 2017.

\bibitem{hackeroneoss}
hackerone.
\newblock Supporting the source: Why hackerone is upgrading its free tools for open source.
\newblock \url{https://www.hackerone.com/vulnerability-management/supporting-source-why-hackerone-upgrading-its-free-tools-open-source}, 2019.

\bibitem{h1report}
HackerOne.
\newblock 7th annual hacker-powered security report.
\newblock \url{https://www.hackerone.com/reports/7th-annual-hacker-powered-security-report}, 2023.

\bibitem{ljharbdeps}
Jordan Harband.
\newblock A querystring parser with nesting support (qs).
\newblock \url{https://github.com/ljharb/qs/network/dependents}, 2015.

\bibitem{repro1}
Ahmed Hassan.
\newblock Sql database error could lead to sql injection with internal path disclosure in froxlor/froxlor.
\newblock \url{https://huntr.com/bounties/4ab24ee2-3ff6-4248-9555-0af3e5f754ec/}, 2023.

\bibitem{bbquote11}
HDVinnie.
\newblock Cross-site request forgery (csrf) in bookstackapp/bookstack.
\newblock \url{https://huntr.com/bounties/114bfbc2-850a-4116-aa07-0d666a9626de/}, 2021.

\bibitem{he2023automating}
Runzhi He, Hao He, Yuxia Zhang, and Minghui Zhou.
\newblock Automating dependency updates in practice: An exploratory study on github dependabot.
\newblock {\em IEEE Transactions on Software Engineering}, 2023.

\bibitem{hunterguide}
Huntr.
\newblock Participation guidelines.
\newblock \url{https://huntr.com/guidelines}, 2020.

\bibitem{huntr_hacktivity_page}
Huntr.
\newblock Hacktivity.
\newblock \url{https://huntr.com/bounties/hacktivity/}, 2021.

\bibitem{rsp1}
Karin Höne and Jan H.~P. Eloff.
\newblock What makes an effective information security policy?
\newblock {\em Network Security}, 2002(6):14--46, 2002.

\bibitem{ibexaadv}
ibexa.
\newblock Ineffective object state limitation and unauthenticated fastly purge.
\newblock \url{https://developers.ibexa.co/security-advisories/ibexa-sa-2022-004-ineffective-object-state-limitation-and-unauthenticated-fastly-purge}, 2022.

\bibitem{imtiaz2022open}
Nasif Imtiaz, Aniqa Khanom, and Laurie Williams.
\newblock Open or sneaky? fast or slow? light or heavy?: Investigating security releases of open source packages.
\newblock {\em IEEE Transactions on Software Engineering}, 49(4):1540--1560, 2022.

\bibitem{bbquote8}
Janette88.
\newblock Use after free in function qf\_get\_curlist in vim/vim.
\newblock \url{https://huntr.com/bounties/fa31a7e0-70a8-471b-bf9c-abb04d3ad38e}, 2022.

\bibitem{cybersecuritydive}
David Jones.
\newblock Cves expected to rise in 2023, as organizations still struggle to patch.
\newblock \url{https://www.cybersecuritydive.com/news/cves-rise-2023-struggle-to-patch/641955}, 2023.

\bibitem{adv1}
Gareth Jones.
\newblock Possible denial of service vulnerability in rack's header parsing.
\newblock \url{https://github.com/advisories/GHSA-c6qg-cjj8-47qp}, 2023.

\bibitem{cve2009gh}
JRuby-OpenSSL.
\newblock jruby-openssl gem for jruby fails to do proper certificate validation.
\newblock \url{https://github.com/advisories/GHSA-xgv7-pqqh-h2w9}, 2009.

\bibitem{anotherbb}
Chau~Minh Khanh.
\newblock Failure to invalidate session after password change in bigbluebutton/greenlight.
\newblock \url{https://huntr.com/bounties/9b341840-fd3f-4a21-839f-ad1fcb422a0e}, 2022.

\bibitem{issuehunt}
IssueHunt K.K.
\newblock Issuehunt.
\newblock \url{https://issuehunt.io/}, 2018.

\bibitem{adquote3}
Ben~Dror Kolin.
\newblock Ibexa user settings are accessible on the front-end for anonymous user.
\newblock \url{https://github.com/advisories/GHSA-r3fg-3r88-6x3f}, 2023.

\bibitem{kula2018developers}
Raula~Gaikovina Kula, Daniel~M German, Ali Ouni, Takashi Ishio, and Katsuro Inoue.
\newblock Do developers update their library dependencies? an empirical study on the impact of security advisories on library migration.
\newblock {\em Empirical Software Engineering}, 23:384--417, 2018.

\bibitem{laszka2018engagement}
Aron Laszka, Mingyi Zhao, Akash Malbari, and Jens Grossklags.
\newblock The rules of engagement for bug bounty programs.
\newblock In Sarah Meiklejohn and Kazue Sako, editors, {\em Financial Cryptography and Data Security}, pages 138--159, Berlin, Heidelberg, 2018. Springer Berlin Heidelberg.

\bibitem{snykdb}
Snyk Limited.
\newblock \url{https://security.snyk.io/}.

\bibitem{chengwei2022demystifying}
Chengwei Liu, Sen Chen, Lingling Fan, Bihuan Chen, Yang Liu, and Xin Peng.
\newblock Demystifying the vulnerability propagation and its evolution via dependency trees in the npm ecosystem.
\newblock In {\em Proceedings of the 44th International Conference on Software Engineering}, ICSE '22, page 672–684, 2022.

\bibitem{luna2019productivity}
Donatello Luna, Luca Allodi, and Marco Cremonini.
\newblock Productivity and patterns of activity in bug bounty programs: Analysis of hackerone and google vulnerability research.
\newblock In {\em Proceedings of the 14th International Conference on Availability, Reliability and Security}, pages 1--10, 2019.

\bibitem{bbrdp1}
Jieyong Ma.
\newblock Buffer over-read in hpjansson/chafa.
\newblock \url{https://huntr.com/bounties/f6b9114b-671d-4948-b946-ffe5c9aeb816/}, 2022.

\bibitem{bbrp2}
Jieyong Ma.
\newblock Heap use after free in function skipwhite in vim/vim.
\newblock \url{https://huntr.com/bounties/1eed7009-db6d-487b-bc41-8f2fd260483f/}, 2022.

\bibitem{maillart2017given}
Thomas Maillart, Mingyi Zhao, Jens Grossklags, and John Chuang.
\newblock Given enough eyeballs, all bugs are shallow? revisiting eric raymond with bug bounty programs.
\newblock {\em Journal of Cybersecurity}, 3(2):81--90, 2017.

\bibitem{privdisclosure}
Sergio Marotco.
\newblock Owasp vulnerability disclosure cheat sheet.
\newblock \url{https://cheatsheetseries.owasp.org/cheatsheets/Vulnerability_Disclosure_Cheat_Sheet.html}, 2022.

\bibitem{vccs}
Andrew Meneely, Harshavardhan Srinivasan, Ayemi Musa, Alberto~Rodríguez Tejeda, Matthew Mokary, and Brian Spates.
\newblock When a patch goes bad: Exploring the properties of vulnerability-contributing commits.
\newblock In {\em 2013 ACM / IEEE International Symposium on Empirical Software Engineering and Measurement}, pages 65--74, 2013.

\bibitem{bbsp1}
Khanh~Chau Minh.
\newblock Cross-site request forgery (csrf) in francoisjacquet/rosariosis.
\newblock \url{https://huntr.com/bounties/158bcf1f-91fb-4398-8b8f-a4bcd4e9ba88/}, 2021.

\bibitem{cveprocess2}
MITRE.
\newblock Cve record lifecycle.
\newblock \url{https://www.cve.org/About/Process}, 1999.

\bibitem{cve20213902}
MITRE.
\newblock Cve-2021-3902.
\newblock \url{https://cve.mitre.org/cgi-bin/cvename.cgi?name=CVE-2021-3902}, 2021.

\bibitem{adquote8}
Torbjørn~Birch Moltu.
\newblock Feature discussion - let's upgrade security on the wallet \#2739.
\newblock \url{https://github.com/web3/web3.js/issues/2739}, 2019.

\bibitem{h1popular}
Phil Muncaster.
\newblock Hackerone exceeds \$300m in bug bounty payments.
\newblock \url{https://www.infosecurity-magazine.com/news/hackerone-exceeds-300m-bug-bounty/}, 2023.

\bibitem{mendio}
Adam Murray.
\newblock National vulnerability database explained.
\newblock \url{https://www.mend.io/blog/the-national-vulnerability-database-explained/}, 2018.

\bibitem{cveprocess}
NIST.
\newblock Cves and the nvd process.
\newblock \url{https://nvd.nist.gov/general/cve-process}, 1999.

\bibitem{nistnvd}
NIST.
\newblock National vulnerability database (nvd).
\newblock \url{https://nvd.nist.gov/}, 1999.

\bibitem{nvddist}
NIST.
\newblock Cvss severity distribution over time.
\newblock \url{https://nvd.nist.gov/general/visualizations/vulnerability-visualizations/cvss-severity-distribution-over-time#CVSSSeverityOverTime}, 2022.

\bibitem{hunter}
Adam Nygate.
\newblock Huntr.
\newblock \url{https://huntr.com}, 2020.

\bibitem{openssfwg}
OpenSSF.
\newblock Openssf working groups.
\newblock \url{https://openssf.org/community/openssf-working-groups/}, 2020.

\bibitem{scmopenssf}
OpenSSF.
\newblock Source code management platform configuration best practices.
\newblock \url{https://best.openssf.org/SCM-BestPractices/}, 2023.

\bibitem{rsp2}
Mikko Siponen{,}~Seppo Pahnila{,} and Adam Mahmood.
\newblock Employees’ adherence to information security policies: An empirical study.
\newblock In {\em New Approaches for Security, Privacy and Trust in Complex Environments.}, pages 133--144. IFIP International Federation for Information Processing, vol 232., 2007.

\bibitem{vccfinder}
Henning Perl, Sergej Dechand, Matthew Smith, Daniel Arp, Fabian Yamaguchi, Konrad Rieck, Sascha Fahl, and Yasemin Acar.
\newblock Vccfinder: Finding potential vulnerabilities in open-source projects to assist code audits.
\newblock In {\em Proceedings of the 22nd ACM SIGSAC Conference on Computer and Communications Security}, page 426–437, New York, NY, USA, 2015.

\bibitem{starbs}
peuch.
\newblock Information leak - github - jms information.
\newblock \url{https://hackerone.com/reports/360811}, 2018.

\bibitem{venturebeat}
Taryn Plumb.
\newblock Github’s octoverse report finds 97 percent of apps use open source software.
\newblock \url{https://venturebeat.com/programming-development/github-releases-open-source-report-octoverse-2022-says-97-of-apps-use-oss}, 2022.

\bibitem{bbsp4}
Bruno~Salvatierra Préntice.
\newblock Arbitrary template creation leading to authenticated rce in hay-kot/mealie.
\newblock \url{https://huntr.com/bounties/3ecd4a78-523e-4f84-a3fd-31a01a68f142/}, 2022.

\bibitem{ghissue1}
Rajat Raghav.
\newblock Security issue: Code execution \#2837.
\newblock \url{https://github.com/johannesjo/super-productivity/issues/2837}, 2023.

\bibitem{raymond1999cathedral}
Eric Raymond.
\newblock The cathedral and the bazaar.
\newblock {\em Knowledge, Technology \& Policy}, 12(3):23--49, 1999.

\bibitem{psfdeps}
Kenneth Reitz.
\newblock Requests.
\newblock \url{https://github.com/psf/requests/network/dependents}, 2019.

\bibitem{ruohonen2018bug}
Jukka Ruohonen and Luca Allodi.
\newblock A bug bounty perspective on the disclosure of web vulnerabilities.
\newblock {\em arXiv preprint arXiv:1805.09850}, 2018.

\bibitem{shafigh2021bugbounty}
Saman Shafigh, Boualem Benatallah, Carlos Rodr\'{\i}guez, and Mortada Al-Banna.
\newblock Why some bug-bounty vulnerability reports are invalid? study of bug-bounty reports and developing an out-of-scope taxonomy model.
\newblock In {\em Proceedings of the 15th ACM / IEEE International Symposium on Empirical Software Engineering and Measurement (ESEM)}, New York, NY, USA, 2021. Association for Computing Machinery.

\bibitem{bbsp2}
Jaylon Simmons.
\newblock Path traversal in misp/misp-maltego.
\newblock \url{https://huntr.com/bounties/0818e9c9-c5fa-4827-a942-8302c96c04ff/}, 2021.

\bibitem{snykadv}
Snyk.
\newblock Regular expression denial of service (redos).
\newblock \url{https://security.snyk.io/vuln/SNYK-JS-SEMVER-3247795}, 2023.

\bibitem{sridhar2021hacking}
Kiran Sridhar and Ming Ng.
\newblock Hacking for good: Leveraging hackerone data to develop an economic model of bug bounties.
\newblock {\em Journal of Cybersecurity}, 7(1), 2021.

\bibitem{darkreading}
Dark~Reading Staff.
\newblock Supply chain attack deploys hundreds of malicious npm modules to steal data.
\newblock \url{https://www.darkreading.com/attacks-breaches/supply-chain-attack-malicious-npm-modules-steal-data}, 2023.

\bibitem{cURL}
Daniel Stenberg.
\newblock curl.
\newblock \url{https://curl.se/}, 1996.

\bibitem{bbquote3}
Bryce Sullivan.
\newblock (almost) arbitary file read on development server in nuxt/nuxt.
\newblock \url{https://huntr.com/bounties/7840cd32-af15-40cb-a148-7ef3dff4a0c2}, 2023.

\bibitem{html2text}
Aaron Swartz.
\newblock html2text.
\newblock \url{https://github.com/aaronsw/html2text}, 2011.

\bibitem{bbrp1}
Mahendra Thanniru.
\newblock Leaking password protected articles content due to improper access control in publify/publify.
\newblock \url{https://huntr.com/bounties/b398e4c9-6cdf-4973-ad86-da796cde221f/}, 2022.

\bibitem{induc2006thomas}
David~R. Thomas.
\newblock A general inductive approach for analyzing qualitative evaluation data.
\newblock {\em American Journal of Evaluation}, 27(2):237--246, 2006.

\bibitem{avastcite}
Kevin Townsend.
\newblock The disclosure of vulnerabilities and the vulnerability of disclosures.
\newblock \url{https://blog.avast.com/the-importance-of-vulnerability-disclosure-avast}, 2020.

\bibitem{utz2022privacy}
Christine Utz, Sabrina Amft, Martin Degeling, Thorsten Holz, Sascha Fahl, and Florian Schaub.
\newblock Privacy rarely considered: Exploring considerations in the adoption of third-party services by websites.
\newblock {\em arXiv preprint arXiv:2203.11387}, 2022.

\bibitem{statistaworldwide}
Lionel~Sujay Vailshery.
\newblock Open source software vulnerabilities worldwide from 2009 to 2020.
\newblock \url{https://www.statista.com/statistics/1245670/worldwide-open-source-software-vulnerabilities}, 2023.

\bibitem{walshe2020empirical}
Thomas Walshe and Andrew~C Simpson.
\newblock An empirical study of bug bounty programs.
\newblock In {\em 2020 IEEE 2nd International Workshop on Intelligent Bug Fixing (IBF)}, pages 35--44. IEEE, 2020.

\bibitem{walshe2022coordinated}
Thomas Walshe and Andrew~C Simpson.
\newblock Coordinated vulnerability disclosure programme effectiveness: Issues and recommendations.
\newblock {\em Computers \& Security}, 123:102936, 2022.

\bibitem{wang2019detecting}
Xinda Wang, Kun Sun, Archer Batcheller, and Sushil Jajodia.
\newblock Detecting "0-day" vulnerability: An empirical study of secret security patch in oss.
\newblock In {\em 2019 49th Annual IEEE/IFIP International Conference on Dependable Systems and Networks (DSN)}, pages 485--492. IEEE, 2019.

\bibitem{disableconvo}
Tim Wojtulewicz.
\newblock Disable "report a security vulnerability" option/button?
\newblock \url{https://github.com/orgs/community/discussions/45567}, 2023.

\bibitem{bbrp3}
Ken Wong.
\newblock Use-after-free in str\_escape in mruby/mruby in mruby/mruby.
\newblock \url{https://huntr.com/bounties/9fcc06d0-08e4-49c8-afda-2cae40946abe/}, 2022.

\bibitem{bbquote1}
wtwver.
\newblock External control of file name or path in hestiacp.
\newblock \url{https://huntr.com/bounties/e0a2c6ff-b4fe-45a2-9d79-1f4dc1b381ab}, 2021.

\bibitem{bbrdp2}
James Yeung.
\newblock Cross-site scripting- stored in pimcore/data-hub.
\newblock \url{https://huntr.com/bounties/708971a6-1e6c-4c51-a411-255caeba51df/}, 2022.

\bibitem{zhao2015empirical}
Mingyi Zhao, Jens Grossklags, and Peng Liu.
\newblock An empirical study of web vulnerability discovery ecosystems.
\newblock In {\em Proceedings of the 22nd ACM SIGSAC Conference on Computer and Communications Security}, pages 1105--1117, 2015.

\bibitem{zhao2017devising}
Mingyi Zhao, Aron Laszka, and Jens Grossklags.
\newblock Devising effective policies for bug-bounty platforms and security vulnerability discovery.
\newblock {\em Journal of Information Policy}, 7:372--418, 2017.

\end{thebibliography}
